\documentclass[aps,rmp,reprint,amsmath,amssymb,graphicx,longbibliography]{revtex4-1}
\usepackage{bm}
\usepackage{amsmath}    
\usepackage{graphicx}   
\usepackage{url}
\newcommand{\ket}[1]{\left|#1\right\rangle}
\DeclareMathOperator{\tr}{tr}

\begin{document}

\title{Quantum Random Number Generators}
\author{Miguel Herrero-Collantes}  
\email{miguel.herrero@incibe.es}  
\affiliation{Instituto Nacional de Ciberseguridad, \mbox{Avenida Jos\'e Aguado, 41,} Edificio INCIBE \mbox{24005, Le\'on, Spain.}\\}

\affiliation{\mbox{EI Telecomunicaci\'on, Department of Signal Theory and Communications, University of Vigo,} Campus Universitario Lagoas-Marcosende, E-36310 Vigo, Spain.}
\author{Juan Carlos Garcia-Escartin}
\email{juagar@tel.uva.es}  
\affiliation{Universidad de Valladolid, Dpto. Teor\'ia de la Se\~{n}al e Ing. Telem\'atica, Paseo Bel\'en n$^o$ 15, 47011 Valladolid, Spain.}

\date{\today}

\begin{abstract}
Random numbers are a fundamental resource in science and engineering with important applications in simulation and cryptography. The inherent randomness at the core of quantum mechanics makes quantum systems a perfect source of entropy. Quantum random number generation is one of the most mature quantum technologies with many alternative generation methods. We discuss the different technologies in quantum random number generation from the early devices based on radioactive decay to the multiple ways to use the quantum states of light to gather entropy from a quantum origin. We also discuss randomness extraction and amplification and the notable possibility of generating trusted random numbers even with untrusted hardware using device independent generation protocols. 
\end{abstract}

\maketitle

\tableofcontents{}

\section{Motivation}
Quantum mechanics offers interesting new protocols in the intersection between computer science, telecommunications, information theory and physics. Results like the protocols for quantum key distribution \cite{BB84,Eke91} and efficient algorithms for problems that are thought or known to be hard for classical computers \cite{EJ96,CvD10} show quantum physics can have a profound impact in the way we think about security, cryptography and computation.

Despite the impressive experimental achievements of the last decades, the current state of technology is still not advanced enough for a full-scale universal quantum computer. Quantum key distribution, on the other hand, has already become an established technology and the first commercial systems have been demonstrated in practical scenarios \cite{PPA09,SFI11}.

Another important well-established quantum technology is quantum random number generation. \emph{Quantum random number generators, QRNGs,} are devices that use quantum mechanical effects to produce random numbers and have applications that range from simulation to cryptography. They are usually simpler than other quantum devices and are mature enough to be applied. QRNGs using different quantum phenomena have gone from the lab to the shelves with at least eight existing commercial products \cite{IDQ, QuT,Pic,QRB,MPD,Com,Qui,HN16} and online servers that provide quantum random numbers on demand \cite{Wal96,Anu,Hum,STK08,Gen}, as well as many patents \cite{DDH02,DH02,TV07,LCL07,BMS08,SZ15,KK01,Kla03,Kla05,RG09,VBG11}. In the last few years there has also been a large number of proposals, experiments, improvements and exciting theoretical results in randomness extraction and randomness certification. 

The aim of this review is to collect the most important proposals for quantum random number generation and give an introduction to the new advanced protocols that use quantum physics to process, certify or otherwise deal with random strings. This paper complements previous surveys on the topics of physical and quantum random number generation \cite{Sti11,SK14} with a focus on QRNGs based on quantum optics.

Section \ref{Applications} gives a brief description of the most important applications of randomness in science and computers. We review the differences between algorithmic methods to produce random looking numbers and physical methods to produce true random numbers and discuss when each method is more appropriate. Due to their importance, we concentrate on applications to simulation and cryptography.

Section \ref{block} describes the main functional elements of quantum random number generators and their roles. In Section \ref{EntEstimation}, we present some mathematical measures of randomness which are particularly useful to analyse the amount of available random bits and the security of quantum random number generators. 

Section \ref{Radioactive} discusses QRNGs based on radioactive decay, which were the first proposed QRNGs and are still in use today. Section \ref{noise} introduces random number generators based on electronic noise and analyses when they can be said to be quantum. 

Section \ref{OQRNG} discusses how optics has modernized QRNGs. Most present-day QRNGs are based on quantum optics and we review the multiple implementations that work with the quantum states of light. 

Section \ref{Other} covers alternative QRNGs based on non-optical quantum phenomena and Section \ref{DeviceIndependent} is centered on those QRNGs whose randomness is backed by quantum mechanics. 

Section \ref{postprocessing} gives a brief tour on the available classical randomness extraction methods and Section \ref{QRandExt} introduces the quantum protocols for randomness expansion and amplification that allow to produce good-quality random outputs from weak randomness sources.

Section \ref{tests} is an introduction to the statistical tests that are usually employed to assess the quality of the final random bit stream.

Finally, in Section \ref{discussion}, we give an overview on the current state of quantum random number generation and the challenges and opportunities for the next generation of quantum devices in the field of randomness.

\section{Random numbers and their applications}
\label{Applications}
Random numbers are an essential resource in science, technology and many aspects of everyday life \cite{Hay01}. Randomness is required to different extents in applications like cryptography, simulation, coordination in computer networks or lotteries. Some applications require a small amount of random numbers and still use manual and mechanical methods to generate randomness, like tossing a coin, throwing a die, spinning a roulette wheel or a drawing a ball from a lottery machine. Here, we will concern ourselves with the generation of random numbers for computers. 

Defining randomness is a deep philosophical problem and we will not attempt to solve it here. In this Section, we give common operational definitions of randomness that fit the different purposes the random numbers must fulfil. For instance, in simulation, a method that generates numbers simulating the statistics of the desired distribution can be considered to be ``random enough'', even if it produces a predictable sequence. 

\subsection{Pseudorandom number generators and true random number generators}
In computing, it is important to distinguish between algorithmically generated numbers that mimic the statistics of random distributions and random numbers generated from unpredictable physical events.

Generating random numbers directly from a computer seems a particularly attractive idea. Methods that produce random numbers from a deterministic algorithm are called {\bf \emph{pseudorandom number generators}}, PRNGs. While it is clear that any algorithmically generated sequence cannot be truly random, for many applications the appearance of randomness is enough\footnote{The famous quote from von Neumann ``Any one who considers arithmetical methods of producing random numbers is, of course, in a state of sin'' is just a way to acknowledge this fact, but also to admit it is an acceptable practice. In the same paper \cite{vNe51} he goes on to comment on some methods to produce pseudorandom sequences.}. 

PRNGs normally start from a small string of bits called the \emph{seed} that is used as the input of a procedure that outputs a long sequence of bits following the statistics of the uniform distribution. In principle, an RNG could produce numbers obeying any random distribution, but the standard practice is trying to provide a uniform distribution, from which we can obtain the most commonly used distributions using well-known transformations \cite{HLD04}. Knuth gives an excellent survey on PRNGs and how to transform uniform random numbers into other types of random quantities in his second book of the series ``The Art of Computing Programming'' \cite{Knu97}.

A large number of PRNGs are based on number theory. Linear congruential generators have been particularly popular since Lehmer introduced them in 1951 \cite{Leh51}. Linear congruential generators produce random numbers from the recursive formula  
\begin{equation}
X_{n+1}=(aX_n+c)\mod m, \hspace{1.5ex}n\geq 0,
\end{equation}
where $X_{i}$ is the $i$th digit in the sequence of random numbers, $m>0$ is the modulus, $0 \leq a < m$ is called the multiplier and $0 \leq c < m$ the increment. The properties of the output depend heavily on the correct choice of these parameters. A poor choice can create an output sequence with a short period. 

Its period is one of the most important properties of any PRNG. The next number in a pseudorandom sequence is determined from the internal state of generator. For a finite memory, the internal state will at some point be the same and the output sequence will begin to repeat itself. PRNGs are chosen to have a large periods so that the repetition does not appear during the intended operation time. 

Apart from congruential linear generators, there is another large family of PRNGs based on linear shift feedback registers, LFSRs, and their generalizations. The most notable generator in this class is the Mersenne Twister \cite{MN98}, which belongs to the family of twisted generalized linear shift feedback registers. The Mersenne Twister has a period which is a Mersenne prime of the form $2^n-1$, for an integer $n$. The most widely used pseudorandom number generator is the MT19937, the standard implementation of the Mersenne Twister with a period $2^{19937}-1$. It is the default generator in many programming languages and popular scientific software. 

L'Ecuyer gives a good review of these and other alternative PRNGs based on different principles \cite{LEc11}.

Pseudorandom numbers have certain advantages that make them popular. They can be much faster than alternative random number generation methods and their results are reproducible. For instance, we can repeat the exact same simulation if we know the seed. However, for many applications, unpredictability is an important requisite. Clearly, a predictable lottery is not acceptable, even if all the resulting numbers are uniformly distributed. Some pseudorandom generators are designed to be unpredictable (see Section \ref{CPRNG}), but, applications that need an output that cannot be guessed usually turn to {\bf \emph{true random number generators}}, TRNGs, if only to renew the seed of a PRNG.

True random number generators measure some unpredictable or, at least, difficult to predict physical process and use the results to create a sequence of random numbers. They either rely on unpredictable values that can be accessed from the software inside the computer or create the sequence in a special-purpose device that feeds it into the operating system.

The process of collecting unpredictable data is usually called \emph{entropy gathering}. Some of the standard entropy sources the operating system can access include data from the sound card, disk access times, the timing of interrupts or user interaction data, like mouse motion or keystrokes, to name a few. The way the Linux operating systems collect entropy and convert it into random bits \cite{GPR06} is an illustrative example of many of the most usual methods. Some authors call these generators that use non-deterministic events, non-physical non-deterministic RNGs \cite{KS08} that stand in contrast to physical TRNGs based in non-deterministic physical effects in electronic circuits or in the result of some physical experiment. Alternatively, there are physical TRNGs based on different principles, such as chaotic systems \cite{SK01,SPK01}, thermal noise in electronic circuits \cite{Mur70,PC00}, free running oscillators \cite{KG04}, or biometric parameters \cite{SWA04} as a few examples.

Some vendors include integrated physical random number generators in their processors. Intel has included in its recent processors a digital RNG based on a metastable latch that, due to thermal noise, suffers jumps in its state at around a 3 GHz rate. This integrated RNG can be directly accessed from a processor instruction, RdRand, \cite{TC11,HKM12}. Similarly, the VIA Technologies Nehemiah processor core includes an on-chip random number generator which is based on a series of oscillators where thermal noise alters the jitter so that the combination of the oscillators' output is random \cite{Cry03}. These integrated RNG include conditioning circuits that process the output to remove biases. 

With an integrated physical random number generator there is always an available source of entropy and we are not limited to resort to other sources of randomness that might not provide fresh entropy in a reliable and steady fashion. For instance, many servers are connected to a limited number of peripherals and do not have access to many random events like mouse motions. These servers can only gather entropy slowly and under severe constraints. 

An integrated physical generator is a convenient addition, but it can also be complemented with the use of external RNGs. This can be a good solution if we do not trust the mechanism in the implementation, the vendor has not released it, or we suspect the chip might have a backdoor either by design or by sabotage \cite{BRP14}.

Quantum random number generators are a particular case of physical TRNGs in which the data is the result of a quantum event. As opposed to other physical systems where uncertainty is a result of an incomplete knowledge of the system, true randomness is an essential part of quantum mechanics as we know it. 

On first sight, physical RNGs seem more desirable that deterministic methods. However, there are inconveniences that have impeded their wider adoption. Some of the problems in physical RNGs are
\begin{enumerate}
\item Limited generation rate. Physical RNG usually produce random numbers at a much smaller rate than software methods. In many cases, there is a fundamental limitation in the rate of change of the sampled physical parameter. If the system is sampled at a high rate, there is not enough time for the system to change and the random numbers are not independent. 
\item It is difficult to give a convincing argument for the randomness of the data. There can be reasonable doubts about the randomness of the chosen physical phenomenon. Many physical random number generators rely on our ignorance to describe a physical process rather than in its intrinsic randomness.
\item Adding an external device is usually inconvenient. 
\item Failures are difficult to detect. If a hardware random number generator fails during operation, it can be difficult to notice. Official recommendations suggest introducing a startup test, a total failure test and an online test to check errors during operation \cite{SK03,KS11}.
\end{enumerate}

The advanced Quantum Random Number Generators that have appeared with the impulse of quantum information research try to solve some of these shortcomings of traditional TRNGs. They offer a solution based on a trusted randomness source and many from the different implementations achieve fast generation rates, normally above the megabit per second, as we will see in the multiple optical implementations described in Section \ref{OQRNG}. This faster rate allows new applications for TRNGs, such as online casinos and Internet gambling, which require a constant stream of random data and cannot use the slower methods of traditional daily or weekly lotteries \cite{IDQ11,Pok16}. 

An important distinction between pseudorandom number generators and physical random number generators is the focus on product or process randomness \cite{Eag05,Cal15}. For pseudorandom number generators we can only evaluate the output strings. We focus on the product of the ultimately deterministic algorithm and we try to determine whether the string has all the properties of a random sequence. In order to determine if we have product randomness our options are limited to checking the output strings and submitting them to certain statistical test (see Section \ref{tests}).

In physical random number generators we concentrate on process randomness. We look for a process that produces a random output and seek to obtain true random numbers from fundamentally random physical phenomena. Here, randomness is usually taken as unpredictability. 

While, properly, classical phenomena can not be considered truly random, in common use, the terms physical and true random number generator are used interchangeably. Usually, it is fine to use an unpredictable physical system as a randomness source. However, there remains a doubt whether the backing physical process is truly random or, at least, presents serious difficulties to be predicted, like a chaotic system, or we simply have a poor model and a better one could destroy the illusion of randomness. Quantum random number generators excel in that aspect: they use very well defined inherently random processes as the source of their bits. 

In the rest of this Section we will consider in some detail how algorithmic and physical random number generation methods are employed in two of the most important families of applications for RNGs, simulation and cryptography. We go through the particular requirements of randomness of each application and discuss the RNGs that are currently used in each case and the dangers of choosing a wrong randomness generation method. We then write about random number generation in fundamental science experiments.

\subsection{Random numbers in simulation}
Random numbers play an essential role in many scientific fields. They are fundamental ingredients in randomized algorithms, which have a wide range of applications in simulation, computing, number theory and other branches of science and engineering \cite{Kar91,MR96}. 

Simplified models of the reality are indispensable tools when we want to predict the behaviour of complex systems that cannot be accurately described with a closed formula or when the computational needs for a full numerical analysis are too high. These models turn to random numbers to incorporate the combined effect of all that is left out. Thus, random number generation is needed in simulations in engineering, network, manufacturing, business and computer science problems \cite{Fis78,BFS87,LK00}. The usual hypothesis is that we can obtain accurate results if we study enough cases chosen uniformly at random. These results, while probabilistic, are usually representative. We need, nevertheless, good random numbers. For instance, in the social sciences it is crucial to have a sound random sampling method to be confident that the study group is a faithful proxy for the whole population that we want to describe \cite{Loh10}. 

A particularly important area is Monte Carlo and Quasi-Monte Carlo methods \cite{MU49,Gen09,Nie78} in which we can find the solution to a complex problem by averaging many random instances. These methods are very effective in solving problems in statistical physics and numerical integration, where they are extensively used. If we sample the state space really at random, the result is likely to be correct, but, due to the high volume of data they require, these algorithms usually get their random numbers from a PRNG.  When correctly done, this is enough. In simulation we only need a generator following the right statistics. However, certain generators that seem reliable under the usual tests (see Section \ref{tests}) have undetected long range correlations that can result in a wrong solution. This is a general problem for congruential generators. In ``Random numbers fall mainly in the planes'' \cite{Mar68} Marsaglia showed that, choosing the right coordinates, consecutive random numbers from multiplicative congruential generators cluster into clear patterns. There are ways to correct this bias \cite{BM07}, but there exist many examples of simulations using faulty PRNG that gave results that, when compared to a known solution, were proved to be wrong, while a different, better PRNG gave the correct answer. There are numerous recorded cases of such failures for the Ising model \cite{KW84,PR85,HCB85,MBH86,FLW92,SW95,OS04} and related problems \cite{Gra93,SHB97,Zif98,HMM10}. Choosing a bad seed during initialization can also result in a correlated output \cite{MWK07}.

Because of these issues, there are authors that have proposed to test PRNGs with the practical problems they are going to solve in addition to the standard statistical tests \cite{Cod94,Cod96, VAK94,VAK95}. For Monte Carlo methods it is also a generally good idea contrasting the results of the same algorithm with different PRNGs, which are unlikely to have the same kind of bias.

True RNG are seldom used for simulation apart from seeding the PRNG. They face several challenges. They are slow when compared to the fastest PRNGs and their results are not easy to reproduce. This is a problem during debugging and replication. The only way to repeat the results of a TRNG is storing the sequence, which can be extremely large for a Monte Carlo run. They also need a fast method to interface with the processor. Anyway, true random number generators are adequate for simulation. While the generation rates of present Quantum RNGs are still a few orders of magnitude below those of good quality PRNGs, they are growing and QRNGs have shown they can be used, at a speed disadvantage, in Monte Carlo simulation \cite{PJL11}. Improvements in the generation speed could make them a viable alternative in certain applications.

\subsection{Random numbers in cryptography}
\label{CPRNG}
Randomness is also a basic cryptographic primitive. Most of modern cryptography follows Kerckhoffs's principle \cite{Ker83} and assumes any cryptographic system can fall into the hands of the adversary and that all the details of the system are perfectly known. Cryptographic system are therefore open and all the security rests in the choice of a secret key. That way, if a channel is compromised, the users just need to change that key. This has many advantages and is generally considered good practice.

In that context, it is of the utmost importance to choose a random key, which usually means choosing an $n$-bit string uniformly at random from all the key space. Similarly, random numbers with sometimes more relaxed randomness requisites are needed in other cryptographic protocols \cite{Gen06}. Random numbers are required in nonces (numbers that must be used only once), in initialization vectors, in sequence numbers \cite{RFC1948}, in salt\footnote{Passwords should not be stored directly as text to prevent further damage if the password file is compromised. The common practice is to store the cryptographic hash of the password string, which, ideally, is a fixed-length bit string that looks random and from which it is unfeasible to recover the original password. However, it is easy to compile a list of the most common passwords and create a list of their hashes. This is called a dictionary attack and it allows an adversary to find the original password from the hashed password list by comparison. One way to hamper this attack is to include a random sequence, called salt, that is hashed together with the password. The salt string is public, but different for every password in the list, making dictionary attacks computationally costly (precomputed universal tables are no longer a valid shortcut).} to avoid dictionary attacks in hashed password lists and in digital signatures, as well as in many interactive protocols \cite{Gol99}.

Quantum cryptography also needs a reliable randomness source. Quantum key distribution is open to attacks if the measurement bases and the states are not chosen in a truly random way, as has been shown for the BB84 protocol \cite{BPP12,LYW15}.

In cryptography it is not enough that the random numbers are uniform. They must also be unpredictable and the generator should limit the damage of any compromised key. There are, depending on authors, at least two new conditions for random numbers to be used in cryptography:
\begin{enumerate}
\item Unpredictability (forward security): an attacker that knows the whole sequence cannot guess the next bit with a probability better than one half. 
\item Backward security: knowledge of a part of the sequence shall not permit an attacker to compute the previous values of the generator with better accuracy than guessing. 
\end{enumerate}
For practical purposes, both requisites of unpredictability can be reduced to polynomial-time unpredictability: that no algorithm can take a subsequence from the generator and guess efficiently (in polynomial time) any previous or following subsequences with better results than total random guessing. This concept is based on Yao's definition of indistinguishable sources \cite{Yao82}.

Most PRNG are not up to the task of generating cryptographically secure random numbers. For instance, the internal state in the Mersenne Twister can be deduced from a long enough output sequence \cite{MN98} and the output of a large type of general congruential generators can be predicted without even knowing the parameters in the generator \cite{Kra90}. 

There are however, established ways to use pseudorandom number generators in cryptographic applications. Algorithmic generators that fulfil the additional criteria are called cryptographically secure pseudorandom number generators CSPRNGs. Two examples based on number theory are the Blum and Micali \cite{BM84} and the Blum Blum Shub generators \cite{BBS86}. We can use Blum Blum Shub as an illustration. The output bits come from the recursive formula
\begin{equation}
X_{i+1}=X_i^2 \mod N
\end{equation}
for $N=pq$ the product of two primes $p$ and $q$ congruent to $3 \mod 4$. $X_i$ is the $i$th number used as the internal state. The algorithm has $N$ and $X_0$ as inputs and the $i$th output bit is the parity of $X_i$ (or, in some variations, a few least significant bits). The initial state $X_0$ should come from a TRNG. This generator has some desirable properties as long as certain common computational complexity assumptions hold. For instance, even if an attacker learnt the internal state $X_i$ at stage $i$, we keep unpredictability to the left (the preceding bits of the binary string are not compromised). Guessing $X_{i-1}$ from $X_i$ is computationally hard unless the quadratic residuosity problem can be solved in polynomial time. Later work showed that breaking Blum Blum Shub is equivalent to factoring \cite{VV85}. This is considered computationally secure in many cryptographic protocols. However, an attacker with a quantum computer that knows $N$ could use Shor's algorithm for integer factorization to break the security of the generator \cite{Sho97}.

There are also variations of the Mersenne Twister intended to make it secure for cryptographic use \cite{MNH05,MSN08}. Other approaches to CSPRNGs use cryptographic protocols such as DES or AES as blocks that transform a string of bits using as their secret key a processed seed from the computer's entropy pool. An example is the random number generation recommendation for banking in the ANSI X9.17 key management standard \cite{ANSIX917}.

There are different standards and recommendations for the cryptographic use of random number generators in key generation \cite{BR12,FIPS1402}
and in financial systems \cite{ANSIX982}, with instructions on how to treat the sources of entropy for seeding PRNGs \cite{TBK16,ISO18031}.  

Cryptographical random number generators, as any critical part in a cryptographic protocol, can be subject to different cryptanalytic attacks \cite{KSW98}. There are also some quantum attacks that offer a moderate advantage with respect to classical strategies \cite{GAL13}.

Certain generators are specifically designed for cryptography and are built to avoid common attacks. An example is the Fortuna pseudorandom number generator that uses multiple sources of entropy to reseed as frequently as possible so that, if the generator is compromised at some time, the previous output remains unguessable \cite{FSK10}. This and similar cryptographic generators are configurable and allow to replace the protocols inside their constituent blocks. 

The design of cryptographically secure RNGs is far from trivial. There are multiple cases of faulty implementations of RNGs that have led to serious vulnerabilities. One common pitfall is the failure to properly seed the generator. Even if the transformation on the seed is secure and cannot be inverted, if there is not enough entropy an attacker can launch a brute force attack and try all the possible seeds. The outputs can then be compared to the output of the generator and the attacker can predict which keys the user has generated. This has happened many times since the early attacks on the SSL keys generated in the Netscape Browser, which used predictable sources like the time of the day or process numbers to seed its generator \cite{GW96,She96}. Similarly, a bug int the OpenSSL library resulted in a seed of limited entropy that used as its only randomness source process identifiers, PID, which have only $2^{15}$ possible values \cite{Ahm08}. The resulting possible keys could be generated by brute force in a few hours. Poor initialization can also weaken the random numbers in operating systems like Windows 2000 \cite{DGP09}. A few more examples of vulnerabilities due to initialization problems or other bad quality random number generators are weak RSA key generation in network devices \cite{HDW12,LHA12}, repeated or guessable keys produced inside smart cards \cite{NES08,BCC13,CHH13} and the predictable random sequences that are used for cryptographic purposes in Android \cite{KHL13,MMS13}.

In this respect, physical RNGs, including QRNGs, can serve as way to seed CSPRNGs, preferably as an additional source of entropy. There are still some important precautions. Certain attacks specifically target TRNGs \cite{ZM97,SCD13} and they can be sensitive to environmental variables \cite{SCC11}. There are already some proposals to test QRNG \cite{WSC15} under the online test of the BSI AIS 20/31 standard \cite{KS11} to make sure they do not fail during operation. As long as these aspects are taken into account, the relatively high rate of QRNGs makes them also a viable option to directly generate keys, probably after some kind of postprocessing. 

In fact quantum key distribution, QKD, \cite{BB84,Eke91,GRT02,SBC09,LCT14} can be seen as nothing more than a very sophisticated distributed secure random number generator that includes a physical method to generate entropy and a randomness amplification algorithm that weeds out the bits that could have been compromised \cite{OHN08}.

In that interpretation, many quantum hacking methods can be considered as attacks to an RNG or to the randomness generation block inside the QKD system \cite{Sti14}. For instance, in detector blinding attacks \cite{LWW10,GLL11}, an attacker can selectively disable the detectors in the receiver and eliminate any randomness in the measurement, determining the result. Similarly, time shift attacks take advantage of different detection efficiencies with time to make measurement in a chosen basis more or less likely introducing a bias \cite{ZFQ08} and attacks based on imperfect beam splitters perform a similar feat by introducing unbalances in the way the quantum states are directed to each measurement configuration \cite{LWH11}.

QKD protocols assume they have access to true randomness and QRNGs are quite adequate for that purpose. We will see they are faster than alternative TRNGs, produce random numbers of good quality and suppose small deviations from the usual configuration of the equipment (they can be built with the same technology and their cost is only a small fraction of the total).

\subsection{Random numbers in fundamental science}
Finally, truly random numbers play a special role in experiments that try to determine the nature of the world. For philosophical reasons, in some proof of principle experiments we need to remove any possible bias when choosing a measurement or when making other decision. To this respect, quantum random number generators stand in a privileged position. Quantum mechanics is the only theory that, according to our understanding, offers true randomness.

This is particularly important in many experiments on the foundations of quantum mechanics, where many of the thought experiments that helped to shape our understanding of the quantum theory have entered the lab and can be tested experimentally \cite{SML14}. Quantum random numbers can also appear in any experiment where we want to be sure there is no hidden bias or that our decisions are independent from previous states of the system. 
Curiously, one of the early quantum random number generators based on radioactive decay, described in Section \ref{Radioactive}, was designed as a way to remove bias in parapsychology  experiments \cite{Sch70,Sch70c}. Later, QRNGs have become part in experiments where randomness is a philosophical necessity. 

Quantum random number generators are a good solution in experiments that test the predictions of the quantum theory. They can be built with equipment similar to that of the experiment or even be integrated into the experimental setup. While we must trust the inherent randomness of quantum effects, they can be instrumental in exploring other aspects of quantum mechanics like complementarity or nonlocality that are not directly dependent on the randomness of quantum measurement. Experimental tests of properties like the wave-particle duality usually require to take random decisions in a short time and quantum random number generators can fulfil that mission. 

Experimental tests of Bell's Inequality \cite{BCP14} need a random choice of basis which can be done with an external QRNG connected to a switch like in the experiments of \cite{WJS98} and \cite{SUK10} that used the QRNG in \cite{JAW00} or with a passive choice, where the quantum randomness comes from separating the paths of the photons in the experiment in a balanced beam splitter \cite{TBG99}, which can be equivalent in the right conditions \cite{GZ99}. 

We also need true randomness for Wheeler's delayed choice experiment in which a photon inside an interferometer can behave like a wave or a particle depending on whether we close the interferometer or not \cite{Whe78}. If the choice is delayed to after the photon is inside the interferometer, the photon must be able to behave both as a wave and a particle\footnote{Indeed, the experiments show the photon can also behave as different combinations in between, with different degrees of visibility and distinguishability.} as the complete setup had not been decided when the photon entered it. From a fundamental point of view, it is crucial that the decision is made after the photon enters the interferometer. We need a fast and truly random number generator. The experiment in \cite{AJW84} uses a single photon from a weak light source with a 50\% probability of firing a detector connected to a switch and the experiments in \cite{JWG07,JWG08} make this decision using a QRNG based on the measurement of the amplified shot noise of white light.

Other experiments include delayed-choice experiments based on entanglement swapping \cite{YS92,ZZH93} after Peres's proposal \cite{Per00} in which whether two photons are entangled or not is decided \emph{after} they have been measured \cite{MZK12} and in quantum erasure experiments that erase path information \cite{MKQ13}, in both cases using the QRNG of \cite{JAW00}.

\section{Block description}
\label{block}
Physical random number generators can be divided into separate blocks with well-defined subtasks. The two most important blocks are the entropy source and the postprocessing stage. The \emph{entropy source} consists of a physical system with some random physical quantity and the measurement equipment that reads these random variables. In digital random number generators we usually need to convert analog measurements into bit strings with the help of analog-to-digital converters. Measurement and quantization are noisy processes and there will be some contamination in what is called the \emph{raw bit string} even if the measured quantity is truly random and free from correlations. The \emph{postprocessing block} takes the raw bits and distills a shorter sequence without correlations. 

The most important phase in postprocessing is randomness extraction. Randomness extractors are functions that transform the bits from the raw sequence into a uniform random sequence at the output with most or all of the randomness available in the input. 

\begin{figure}[ht!]
\includegraphics{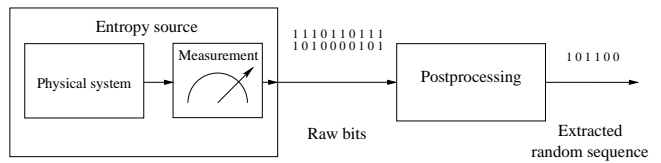}
\caption{\label{BlockDiagram} Block diagram of a typical physical random number generator. A measurement system registers an unpredictable magnitude from a well-characterized physical system and converts the results into a binary raw bit sequence, which can still show some bias. The postprocessing stage extracts a smaller, ideally bias-free, random sequence assuming some bound to the amount of randomness of the raw sequence. The estimation usually comes from a thorough analysis of the original random physical system and the measurement errors.}
\end{figure}

Figure \ref{BlockDiagram} shows the block diagram of a typical physical random number generator. The exact parts vary from device to device. For instance, some physical random number generators are designed to produce raw sequences with negligible bias and forgo the postprocessing phase. There is a delicate balance in choosing an adequate postprocessing system. More involved randomness extraction methods usually allow to minimize the amount of random bits that are thrown away, but are slower. The overall bit rate depends on whether the increased production of bits compensates or not for the slower processing circuit or if it is justified to use a faster but more complex hardware to remove biases from the raw bit sequence.

In this review, we concentrate on the different quantum systems that can work as an entropy source. Section \ref{Radioactive} describes measurements of radioactive decay. Section \ref{OQRNG} explains the many possible sources of entropy available in quantum optics. Section \ref{Other} discusses alternative quantum systems that do not use light. 

Section \ref{postprocessing} gives a brief review on some classical postprocessing algorithms used to remove existing biases and Section \ref{QRandExt} introduces different quantum protocols that can be combined with imperfect randomness sources to obtain uniform output strings.

Before describing the particular systems from which quantum random number generators obtain randomness, in Section \ref{EntEstimation} we comment the most common ways to measure entropy and the contexts in which each entropy measure can be applied. Different authors choose different criteria that will be mentioned when we describe the corresponding quantum random number generator.  

In certain quantum random number generators, like device independent generators (Section \ref{DevInd}), the physical measurement process and randomness estimation and extraction are intimately linked and we discuss them together. 

\section{Entropy estimation}
\label{EntEstimation}
Entropy in its many forms offers a convenient way to measure randomness. The different entropies give a mathematical measure for surprise (how unexpected a value is). We express entropy in bits, in the information theory sense, which is closely related to the concept of thermodynamic entropy but takes it to a more natural formulation for information processing and communications.  

A simple interesting measure is Shannon entropy \cite{Sha48}. For a random variable $X$ with a probability distribution $P_X$ so that $P_X(x)$ is the probability of getting the outcome $x$ from a discrete set $\mathcal{A}$ (an alphabet) with $N$ possible values for $x$, the Shannon entropy of $X$, $H(X)$, is defined as
\begin{equation}
H(X)=-\sum_{x\in \mathcal{A}} P_{X}(x) \log_2 P_X(x).
\end{equation}
Shannon entropy gives the average number of bits of information we can extract from a single outcome. For an alphabet of cardinality $N=|\mathcal{A}|$ and a uniform probability distribution, all the results are equally likely and we need $\log_2 N$ bits to describe them. We can imagine we place all the possible outcomes in a table and assign a $\log_2 N$-bit string to each of them. In a uniform random process all the outcomes are equally ``surprising'' and we need to use all the bits. Less surprising distributions where some results are more likely than others would need, on average, less bits to be described. Table \ref{EntropyTable} shows an example of bit representations for the results of throwing a balanced and an unbalanced four-sided die (a tetrahedron). 

Shannon entropy offers a rough estimation of randomness. Ideally, we would like to generate an almost uniform distribution with a Shannon entropy as close to $\log_2 N$ as possible. A higher Shannon entropy means we have a distribution closer to uniform and that we can extract more random bits from the process, but there are other entropy measures that can give us a more useful figure when deciding how to use a randomness extractor to make the most efficient use of the available randomness, as described in Section \ref{postprocessing}.

\begin{table}
\begin{tabular}{|c|c|c|}
\hline
\multicolumn{3}{|c|}{Fair die}\\
\hline
\phantom{a}$x$\phantom{a} & \phantom{a}$P(x)$\phantom{a} & Sequence\\
\hline
1 & $1/4$ & 00\\
2 & $1/4$ & 01\\
3 & $1/4$ & 10\\
4 & $1/4$ & 11\\
\hline
\end{tabular}
\begin{tabular}{c}
\phantom{aa}\\
\end{tabular}
\begin{tabular}{|c|c|c|}
\hline
\multicolumn{3}{|c|}{Loaded die}\\
\hline
\phantom{a}$x$\phantom{a} & \phantom{a}$P(x)$\phantom{a} & Sequence\\
\hline
1 & $1/2$ & 0\\
2 & $1/4$ & 10\\
3 & $1/8$ & 110\\
4 & $1/8$ & 111\\
\hline
\end{tabular}
\caption{Entropy calculation example for a fair and a loaded four-sided die. For each possible outcome of a throw (first column) there is an associated probability shown in the second column. The third column shows a possible way to assign a bit sequence to each outcome. For a balanced die (left table) we have two bits of entropy $H(X)=-4(1/4)\log_2(1/4) =2$. For a loaded die (right table), we have an entropy $H(X)=-(1/2)\log_2(1/2)-(1/4)\log_2(1/4)-2(1/8)\log_2(1/8)=1.75$. For the given encoding, we can check we need an average of $1(1/2)+2(1/4)+2\cdot 3 (1/8)=1.75$ bits to describe the result.}
\label{EntropyTable}
\end{table}

An interesting generalization of Shannon entropy is the family of R\'enyi entropies \cite{Ren61}. The R\'enyi entropy of order $\alpha$ is defined as 
\begin{equation}
H_{\alpha}(X)=\frac{1}{1-\alpha}\log_2 \sum_{x\in \mathcal{A}}P_X(x)^\alpha.
\end{equation}
Shannon entropy corresponds to the R\'enyi entropy in the limit $\alpha\to 1$. For any distribution, R\'enyi entropies obey the inequality
\begin{equation}
H_{\alpha}(X)\geq H_{\beta}(X)
\end{equation}
for $\alpha\leq \beta$. Entropies of a different orders appear in many security proofs and randomness bounds \cite{Cac97}.

A particularly useful quantity is the min-entropy $H_{\infty}(X)$, which comes from taking the R\'enyi entropy when $\alpha\to \infty$. Alternatively, it can be defined as
\begin{equation}
\label{minent}
H_{\infty}=-\log_2 \left(\max_{x\in \mathcal{A}} P_X(x) \right)
\end{equation} 
where we take the logarithm of the probability of the most likely outcome. The min-entropy gives a lower, worst-case bound to all the R\'enyi entropies. $2^{-H_\infty(X)}$ corresponds to the probability of guessing at the first attempt the outcome from measuring a random variable $X$ with a known distribution. The optimal strategy is guessing the result is the most likely one. In the example given in Table \ref{EntropyTable}, for the uniform distribution the min-entropy is $2$, but for the loaded die we have a value $-\log_2(1/2)=1$. If we guess an outcome $1$ we are right half of the time. 

In a distribution with min-entropy $k$, every possible outcome $x$ has a bounded probability $P_X(x)\leq 2^{-k}$. Any probability distribution of min-entropy $k$ can be written as a convex combination of distributions that are uniform for $k$ bits. This gives an important interpretation of min-entropy as the number of uniform bits that can be extracted from a given distribution. Intuitively, if no single string is too likely, for every random outcome we can extract about $k$ bits of ``surprise'', but no more \cite{CG88,Zuc90}.

There are explicit constructions, like Trevisan's extractor \cite{Tre01} and derived functions \cite{Sha02}, that can give almost $k$ bits with a probability distribution as close to uniform as desired, provided there are some ancillary random bits of good quality. There are different kinds of randomness extractors (see Section \ref{postprocessing}) in which min-entropy or derived quantities offer an upper bound on the number of available random bits. 

R\'enyi entropies, including Shannon entropy and min-entropy, can be generalized to study joint distributions where part of the system is in the power of a legitimate user $A$ and part of the system, which can be correlated to the first part, is in the possession of an eavesdropper $B$. In random number generation, the most useful quantity is conditional min-entropy. In the most general case, we can include distributions that come from quantum systems if we consider the density matrix $\rho_{AB}$ of a state in the joint Hilbert space $\mathcal{H}_{AB}=\mathcal{H}_{A}\otimes\mathcal{H}_{B}$ with a subspace that is restricted to $A$, $\mathcal{H}_{A}$, and a subspace only $B$ can access, $\mathcal{H}_{B}$. The conditional min-entropy of $\rho_{AB}$ related to a reduced density state $\sigma_B$ in $\mathcal{H}_{B}$ is defined as
\begin{equation}
H_{\infty}(A|B)_{\rho}= \sup_{\sigma_B} (-\log_2{\lambda}),
\end{equation}
where $\lambda$ is the smallest real number for which
\begin{equation}
\lambda I_{A}\otimes \sigma_B-\rho_{AB}
\end{equation}
is nonnegative \cite{Ren05} when $I_A$ is the identity matrix corresponding to $\mathcal{H}_A$ and we maximize over the density matrices $\sigma_B$ with trace 1 describing the subsystem in $\mathcal{H}_B$. Conditional min-entropy gives how much information about the results of a measurement by $A$ can be inferred from measurements on $B$ alone. For classical distributions, $2^{-H_{\infty}(A|B)_{\rho}}$ gives the probability of guessing the outcomes of $A$ from our knowledge of $B$ using the optimal strategy \cite{KRS09}. If there is no side information (the systems of $A$ and $B$ are uncorrelated), we recover the definition and interpretation of the min-entropy in Equation (\ref{minent}).

When considering randomness extractors, it is also interesting to speak of the smooth min-entropy
\begin{equation}
H^{\epsilon}_{\infty}(A|B)_{\rho}= \sup_{\tilde{\rho}}H_\infty(A|B)_{\tilde{\rho}}
\end{equation}
with a supremum taken over all the nonnegative operators $\tilde{\rho}_{AB}$ of trace 1 that are close to $\rho_{AB}$ in the sense that $||\tilde{\rho}_{AB}-\rho_{AB}||\leq \epsilon$ for the $L_1$-norm $||A||=\tr\sqrt{A^\dag A}$ \cite{KR11}. 

Instead of giving asymptotic parameters, like traditional entropies, smooth entropies give results valid for a single sample of a distribution. In random number generators, smooth min-entropy is useful as an estimator of the amount of random bits we can extract from a randomness source that might be correlated with an external attacker. Smooth min-entropy gives a tight bound on the length of the bits that a randomness extractor can produce from a given joint distribution and still give an output as close to uniform as desired and uncorrelated to any external system \cite{Ren05,KRS09}.

For a general unknown source, estimating the min-entropy is far from trivial. The problem is intractable for any reasonable sampling circuit with limited size \cite{Wat16,LP02}. We can only determine min-entropy from measurement inefficiently. If our randomness source is stable and faraway bits are independent, this cost can be paid just once during characterization. Normally, physical random number generators use conservative, worst-case bounds for the min-entropy based on a deep analysis of the physical origin of the randomness and there are standardized methods for online estimation \cite{TBK16}. In that respect, quantum random number generators offer a clear advantage: their source of randomness is usually a very well defined quantum phenomenon. Quantum theory gives very accurate predictions. When compared to other random number generators that gather noise from complex processes like atmospheric noise, quantum random number generators have the virtue of a precise description of the randomness source which can be used to derive limits to the available min-entropy, even accounting for additional classical noise or the presence of eavesdroppers. 

Nevertheless, even for these well-characterized quantum randomness sources, hidden correlations remain a challenge. There might be memory effects or correlations between consecutive runs of the quantum experiment that gives our random numbers and we must take due care to ensure independence and the lack of any experimental bias. 

\section{Quantum Random Number Generators based on radioactive decay}
\label{Radioactive}
\subsection{The first quantum random number generators}
With the rise of computer simulation during the second half of the 20th century, there was a growing need for electronic random number generators \cite{HD62}. At that time, it was common to find tables of random numbers. The most famous of such compilations is probably the book ``A million random digits with 100,000 normal deviates'' from the RAND Corporation \cite{RAN55}. The numbers in the book were generated using an electronic roulette wheel and were available in punched card format to allow easy interfacing with computers. There also appeared many electronic random number generators  designed to be connected to computers or output devices like teleprinters \cite{Sow72}.

It was only natural for some researchers to turn to the intrinsic source of randomness of quantum phenomena \cite{II56,Man61,Sch70,Vin70}. Radioactive decay was a particularly accessible source of true randomness. Geiger-M\"uller tubes were already sensitive enough to capture and amplify $\alpha$, $\beta$ and $\gamma$ radiation and reliable, well-characterized radioactive samples were available. For simplicity, most radioactivity-based quantum random number generators were based on the detection of $\beta$ radiation (emitted electrons). 

In a Geiger-M\"uller, GM, detector a single particle produces an ionization event that is amplified in a Townsend avalanche \cite{Fri49}. The result is a device that, when correctly configured, produces a pulse for each detected particle. The probability of any given atom to decay in a time interval $(t,t+dt)$ is given by an exponential random variable so that $P(t)dt=\lambda_m e^{-\lambda_m t} dt$ for a material with a decay constant $\lambda_m$. If the sample retains many of its original atoms (we are in times much smaller than the half-life) and the sample-detector system undergoes practically no change during our time interval (the position of the sample is constant, the gas in the GM tube does not become contaminated\ldots), the time between detected pulses is also an exponential random variable. The times are independent from previous results and the number of pulses that arrive in a fixed time period follows a Poisson distribution. The exact rate depends on many factors, but it can be determined experimentally and we can be satisfied that the pulses arrive at independent times \cite{SSS99}. The probability of finding $m$ pulses in an observation period of $T$ seconds is $P_m(T)=\frac{(\lambda T)^m}{m!}e^{-\lambda T}$, where $\lambda$ gives the mean number of pulses we detect in one second for our source and corresponds to the parameter of the exponential distribution. 

The QRNGs we describe in this Section are the forerunners of the present day optical QRNGs we will see in Section \ref{OQRNG} that use similar concepts and circuits, but replace the radioactive source and the GM counter with photon sources and detectors.

The first QRNGs based on radioactive decay share many common elements. Most use digital counters to convert the pulses from the detector into random digits. A digital counter increases its output value by 1 when it receives a pulse at its input and can be reset to start the count from 0. Another key element is timing with a digital clock. These QRNGs can be best explained if we speak in terms of fast and slow clocks to describe clocks with a frequency $\nu$ that is significantly greater or smaller than the mean rate of detection. A fast clock, with $\nu>\lambda$, generates many pulses between Geiger counts and when a slow clock, with $\nu<\lambda$, produces a pulse, there has been enough time to have registered many counts in the GM detector. 

\begin{figure}[ht!]
\includegraphics{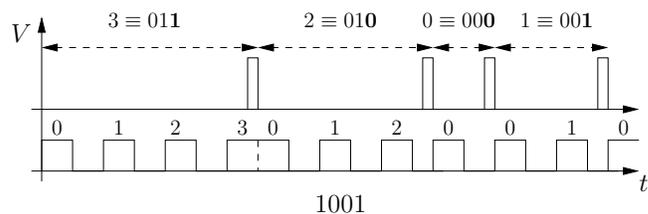}
\caption{\label{CounterFastClock} Fast clock method: A fast clock (down) is used to increase a counter. Whenever a detection is made (up), the counter is read and reset, generating one random number.}
\end{figure}

With these elements, the randomness in the time of arrival can be converted into random digits in a few different ways. The generators of Isida and Ikeda \cite{II56} and Vincent \cite{Vin70} use a counter driven by a fast clock that is read and then reset to zero every time we get a count on the detector. The value of the counter at the moment of the detection is used to produce the random number. Figure \ref{CounterFastClock} gives a graphical description of the method. 
The distribution of values is not uniform and some correction is necessary. If we are producing decimal digits, we can take the least significant figure \cite{II56}. The equivalent method for binary sequences is looking at the parity of the value of the counter, checking if the number of counted pulses is even or odd \cite{Vin70}. This kind of correction draws from previous results for true random number generators that face similar problems \cite{Tho59}. 

\begin{figure}[ht!]
\includegraphics{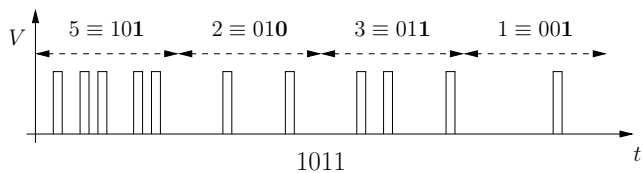}
\caption{\label{CounterSlowClock}Slow clock method:  The Geiger detector is read at fixed intervals, generating a random number that equals to the number of detections during the period.}
\end{figure}

A second option is to use a slow clock to determine when to read the counter. In the generator of Schmidt \cite{Sch70}, the pulses from the GM detector increase the value of a counter. When the slow clock produces a new pulse, the value of the counter is used as a random digit and the count starts again from 0. The output corresponds to the number of particle counts in each clock period. We restrict to a counter that generates values from 0 to $M-1$, a modulo $M$ counter. When $M=2$ we have a binary random number generator. The distribution of the sampled digits is not uniform, but if we take the modulo $M$ addition of multiple outputs, we can obtain a distribution with as small a bias as desired. This is called ``contraction'' and is discussed in detail in Schmidt's paper \cite{Sch70}. Figure \ref{CounterSlowClock} shows an example of this generation method.  

Radioactive decay has also been used to generate white noise for analog computers \cite{Goo54,Man61,How61}. Random noise generation was important, among others, in the analog calculations in airplane design simulations. It also has applications as a test signal and, generally, in communications and simulation problems where a broadband signal is necessary \cite{Gup75}. In this case, the pulses from the GM detector trigger a change of state in a voltage signal. Whenever a particle is detected the signal goes from high to low voltage or from low to high. The resulting random signal is called random telegraph noise \cite{Ric44}. In this case we do not want a binary signal, but Gaussian noise. Instead of sampling, the signal is directed to a low pass filter to complete the noise generator.

\subsection{Evolution}
\begin{figure}[ht!]
\includegraphics{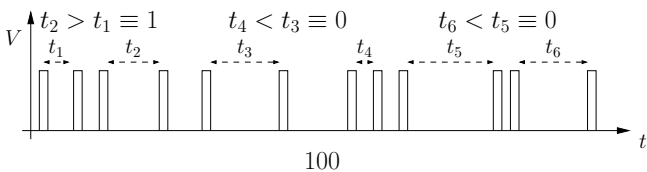}
\caption{\label{TimeDiff} Time difference method: This method compares the time between two events in the Geiger detector. If $t_i<t_{i+1}$ then a bit with value one is generated. Otherwise, the bit generated will be zero.}
\end{figure}

After the initial proposals, there have been different refinements to the basic concept. QRNGs based on radioactive decay are still popular. A good example is the web-based random number server HotBits \cite{Wal96} that has been working since 1996. In the HotBits generator, the random times of arrival of the radiation to the Geiger counter give pairs of intervals of random length. The time between two consecutive pulses is stored as $t_1$ and compared to the time between the next two pulses $t_2$. The random bits come from comparing the times. If $t_1>t_2$ we output a 0 bit and if $t_1<t_2$ we output a 1. The generator reverses the criterion for 0 and 1 for every time pair in order to compensate for small systematic biases that might favour slightly unbalanced intervals. This provides a crude correction for small problems like, for instance, the loss of radioactive material due to radioactive disintegration that makes the second interval shorter on average by a very short time. Figure \ref{TimeDiff} gives a graphical description of the method.  

Some modern proposals replace Geiger counters with semiconductor detectors. Semiconductor devices such as PIN photodiodes can also capture the radiation from radioactive decay \cite{Kno10,Lut07}. Semiconductor detectors are convenient, as they do not require the same high voltage as Geiger tubes. The resulting signal is weaker than that of GM counters, but there are low noise amplifiers that can produce output pulses of a few volts of amplitude. While they can have different sensitivities and need calibration, for the generation of random numbers the important property is not as much determining the actual rate of the particles coming out of the source as it is registering random events. 

Using off-the-shelf semiconductor devices can simplify the design of random number generators. One example of such generators is given by \cite{ANR05} with a variation of the time interval method. Instead of comparing the time between pulses, the system reads a fast clock every time a pulse arrives. If the clock is in a high state (in the high voltage level of the clock cycle) at the moment of arrival the generator outputs a 1. If it is low it outputs a 0. For a good time resolution, the least significant bit of the digitized time should be random and there is no need for postcorrection. 

Two other proposals for QRNGs that use semiconductor detectors with radioactive decay appear in \cite{DLR10}. The first proposal tries to address the problem that in QRNG we have access to an exponential random variable, the time of arrival, or a Poisson random variable, the number of pulses in a fixed time interval. But, in many occasions, RNGs are required to produce uniform random numbers. An exponential random variable of parameter $\lambda$ can be converted to a uniform random variable if we compute:
\begin{equation}
U=e^{-\lambda E},
\end{equation}
where $U$ is the uniform distribution and $E$ the exponential distribution. The first proposal of \cite{DLR10} addresses this with an RC circuit. They use a semiconductor detector whose output pulses trigger the fast discharge of a capacitor. The voltage at the RC circuit when a pulse arrives is the output variable. This approach has several limitations. It needs specialized hardware to transform the voltage to the output and has problems with noise. For that reason there is an alternative proposal with an approach similar to \cite{II56,Vin70}, where a fast clock $\nu \gg \lambda$ drives an $N$-bit counter which is read when a pulse arrives. Here, the clock is supposed to be fast enough to guarantee the samples are uniform in the $2^N$ values. 

\subsection{Limitations}
While QRNGs based on radioactive decay are a good way to obtain high quality true random numbers, they have some drawbacks that limit their practical use. An important barrier is the low bit rate they can achieve, usually below a few hundred kilobits per second. 

The first problem is the need for a radioactive source. In principle, all decay-based QRNGs could work on background radiation. Unless it is isolated, a detector will count stray cosmic rays, radiation from radium, thorium or other radioactive materials in the Earth's crust or particles from radon on air. However, natural activity rarely produces enough particles to cause more that a few counts per second. This poses a fundamental problem for the widespread use of radioactive decay QRNGs. In order to achieve a fast rate, the QRNG needs a highly radioactive source. The reviewed generators used Cobalt-60 \cite{II56}, Strontium-90 \cite{Sch70}, Caesium-137 \cite{Wal96}, Americium-241 \cite{ANR05} or Nickel-63 \cite{DLR10}. This is highly inconvenient and requires improved safety measures. While $\alpha$ sources like Americium are easier to isolate and are common in smoke alarms, the additional precautions prevent straightforward computer integration and this approach works well only for dedicated isolated servers like HotBits \cite{Wal96}. 

A second limitation to the generated bit rate is the \emph{dead time} of the detectors. In Geiger counters the avalanche that amplifies each count ionizes the gas inside the GM tube. The avalanche stops when the positive ions surround the cathode inside the tube. These ions prevent further avalanches until they have returned to their normal state \cite{Fri49}. The dead time is the minimum time for the GM tube to recover its full detection capability and can go from tens of nanoseconds to a few microseconds. This limits the count rate to the MHz range. Semiconductor detectors also need to replenish the carriers after each detection and have dead times in the microsecond range. 

Dead time and other sources of non-uniformity need to be corrected when generating random bits. Vincent describes some important cautions in a follow-up paper \cite{Vin71} to his original generator proposal. In general, the quality of the generated bits will be good and, when there is some residual bias, there exist simple postprocessing methods to recover a random output.

A final problem specific to semiconductor detectors is the damage they suffer from radiation. Geiger tubes also degrade with time, but the effect of radiation on them has been extensively studied, while semiconductors used specifically for radiation detection are relatively new. As long as the damage gives a progressive and slow reduction in efficiency, the output would retain randomness, but more studies on the long term behaviour of these detectors are needed.
 
Despite these constraints, radioactive decay is a suitable source of randomness for low speed applications. It can, for instance, be used to provide entropy for the seed of pseudorandom number generators. For more demanding systems that require high bit rates or when we would like to avoid radioactive sources, the recent optical QRNGs described in Section \ref{OQRNG} are good substitutes.

\section{Random Number Generators based on noise}
\label{noise}
\begin{figure}[ht!]
\includegraphics{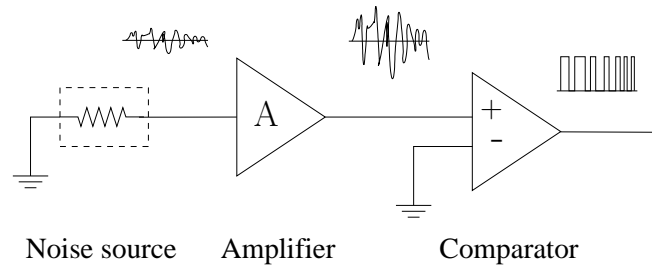}
\caption{\label{AmplifiedNoise}Conceptual representation of a typical noise-based random number generator. The voltage coming from a source of white noise is amplified and compared to a threshold in a comparator to produce a digital signal with random transition times. This signal can be sampled or processed later to give a random bit sequence.}
\end{figure}

Noise in electronic circuits is one of the preferred sources of entropy in classical physical random number generators. Noise appears as an unwanted effect in electronic systems of all kinds and it is readily available. A typical random number generator using noise is shown in Figure \ref{AmplifiedNoise}.

The noise source is represented as a resistor, but other elements can take its place. A Zener diode operated in the reverse breakdown region is another popular choice. In this scheme, voltage fluctuations due to noise are amplified and compared to a threshold to generate random bits. For a threshold of 0 volts, we can sample the amplified noise periodically and assign a 0 if we find a negative voltage and a 1 to a positive voltage. 

If, instead of sampling, we generate a pulse every time the voltage from a white noise source crosses the threshold, the output will be a series of pulses with times of arrival that correspond to a Poisson distribution and we can use any of the methods described in Section \ref{Radioactive} to produce random sequences. The electronic noise circuit replaces the Geiger counter in an otherwise unchanged system. In fact, many proposals for QRNG based on radioactive decay discuss both methods in parallel \cite{Vin70,Gud85}.

There are multiple examples of true random number generators based on this electronic noise like those in \cite{HCD97,PC00} to name a few.  

Noise in those systems comes fundamentally from two sources, shot, or Schottky, noise \cite{Sch18} and thermal, or Johnson-Nyquist, noise \cite{Joh28,Nyq28}, with flicker noise contributing sometimes at low frequencies. Shot noise generates from quantum effects due to the granularity of the current. Currents are formed by discrete carriers and show quantum fluctuations. Thermal noise comes from thermal agitation of the carriers and is produced by statistical motion that depends on the temperature. In practice, both noises tend to appear side by side and are difficult to isolate. In many cases the frontier between shot and thermal fluctuations is blurry \cite{Lan93}. 

In this review, we will not discuss in detail random number generators based on electronic noise. While electronic noise coming from shot fluctuations can be rightfully said to be quantum \cite{RPH98}, it is usually not well characterized and separated from thermal noise, it is subject to many environmental fluctuations and can show memory effects \cite{Sti11}. Somewhat arbitrarily, we choose to concentrate on generators where the quantum effects are well isolated and we have a higher degree of control. Unless there is some interesting effect, we will not discuss true random number generators where quantum noise is only an unquantified part of the total available randomness.

There are a few interesting exceptions. Certain commercial quantum random number generators use electronic noise in semiconductors. For Comscire's QRNG there is a detailed estimation of the quantum entropy gathered from shot noise in MOS transistors \cite{Com}. Likewise, under the right conditions, Zener diodes can be operated in a regime where quantum shot noise dominates \cite{Som75,Sti04}.

\section{Optical Quantum Random Number Generators}
\label{OQRNG}
Most of the existing QRNGs are based on quantum optics. The inherent randomness in many parameters of the quantum states of light allows for a rich choice of implementations. Light from lasers, light emitting diodes or single photon sources is a convenient and affordable substitute for radioactive material as a source of quantum randomness and there are many available detectors. In this section, we study some of the most common ways to harness quantum light to produce random bits.  

First, we give an overview of the concepts of quantum optics that appear in the generators. Then, we propose a classification of optical quantum random number generators, OQRNGs, based on the generation mechanism. Table \ref{SumTable} gives a summary of the covered optical generators with some representative examples, the typical bit rates and the limitations of each kind of generator.

\begin{table*}
\begin{tabular}{|c|c|c|c|c|}
\hline 
Type (Section)& Physical principle & Representative examples& Rate (order) & Challenges \\
\hline
Branching path (\ref{branching})&\begin{tabular}{@{}c@{}} Path superposition \\ +\\ measurement \end{tabular}  & \cite{JAW00}&  Mbps & \begin{tabular}{@{}c@{}}- Unbalanced detectors. \\ - Detector dead time.\end{tabular}  \\
\hline
Time of arrival (\ref{timearrival})& Time of arrival statistics & \begin{tabular}{@{}c@{}}\cite{SR07} \\ \cite{WJA09}\\\cite{WLB11}\end{tabular} &  Mbps &  \begin{tabular}{@{}c@{}}- Time precision. \\ - Detector dead time.\end{tabular}   \\
\hline
Photon counting (\ref{photoncounting})& Photon number statistics &  \begin{tabular}{@{}c@{}}\cite{FWN10} \\ \cite{RWL11}\end{tabular} &  Mbps &  \begin{tabular}{@{}c@{}}- Photon resolving capability. \\ - Detector dead time.\end{tabular}  \\
\hline
Attenuated pulse (\ref{attpuls})& \begin{tabular}{@{}c@{}} Binary measurement \\ of coherent states \end{tabular}  &  \cite{WG09b} &  Mbps &  \begin{tabular}{@{}c@{}}- Source instability. \\ - Detector dead time.\end{tabular}  \\
\hline
Vacuum fluctuations (\ref{vacuum})& Shot noise measurement &  \begin{tabular}{@{}c@{}}\cite{GWS10} \\ \cite{STZ10}\\ \cite{SAL11}\end{tabular} &  Mbps-Gbps & \begin{tabular}{@{}c@{}}- Classical noise. \\ - Postprocessing.\end{tabular}   \\
\hline
Phase noise (\ref{phasenoise})& Laser phase noise &  \begin{tabular}{@{}c@{}}\cite{GTL10} \\ \cite{QCL10}\\ \cite{JCS11}\end{tabular} &  Gbps &  \begin{tabular}{@{}c@{}}- Phase drift. \\ - Pulse repetition rate.\end{tabular}  \\
\hline
\begin{tabular}{@{}c@{}}Amplified Spontaneous  \\ Emission, ASE (\ref{ASE}) \end{tabular} & \begin{tabular}{@{}c@{}} Amplitude fluctuations in \\ ASE noise \end{tabular}   &  \begin{tabular}{@{}c@{}}\cite{WSL10} \\ \cite{APD12}\end{tabular} &  Gbps &  \begin{tabular}{@{}c@{}}- Sampling/digitization. \\ - Postprocessing.\end{tabular}  \\
\hline
Raman Scattering (\ref{Raman})&  \begin{tabular}{@{}c@{}}Interaction with \\ phonon fluctuations\end{tabular}  &  \begin{tabular}{@{}c@{}}\cite{BML11} \\ \cite{CCX15}\end{tabular} & kbps-Mbps &  \begin{tabular}{@{}c@{}}- Raman gain. (Stimulated) \\ - Detector dead time.\\ (Spontaneous)\end{tabular}  \\
\hline
\begin{tabular}{@{}c@{}}Optical Parametric \\ Oscillators, OPOs (\ref{OPO}) \end{tabular} &  \begin{tabular}{@{}c@{}}Bistability in optical \\ parametric oscillators\end{tabular}  &  \begin{tabular}{@{}c@{}}\cite{MLV11} \\ \cite{MLV12}\end{tabular} &  kbps &  \begin{tabular}{@{}c@{}}- Cavity decay time. \\ - Pump repetition rate.\end{tabular}   \\
\hline\end{tabular}
\caption{Summary of the optical methods for quantum random number generation. The table gives the section where we describe the details of each implementation, the principle of operation, a few representative examples, the order of magnitude of the typical bit rates of each generator and a list of the most important limitations.}
\label{SumTable}
\end{table*}

\subsection{Quantum optics in random number generators}
The optical field can be described at the quantum level in terms of photons \cite{KS68,Lou00}. From the many possible families of quantum states, Fock states and coherent states give the most relevant description of the quantum states of light in random number generators. Fock states, or number states, are described as states $\ket{n}$ that contain $n$ photons sharing a mode (they have the same frequency, polarization, temporal profile and a common path). Coherent states, which share many properties with classical light, can be written as a superposition of number states 
\begin{equation}
\label{coherent}
\ket{\alpha}=e^{-\frac{|\alpha|^2}{2}}\sum_{n=0}^{\infty} \frac{\alpha^n}{\sqrt{n!}}\ket{n}
\end{equation}
where $\alpha$ is a complex number. The amplitude $|\alpha|^2$ corresponds to the mean photon number of the state. Weak laser light is an excellent  approximation to a coherent state. We can also use the coherent states from a laser to produce a proxy for single photon states by choosing a low enough intensity, as it usual, for instance, in quantum key distribution with typical values of $\alpha$ around 0.1.

In many applications we are only interested in producing uncorrelated single photons. In that case, attenuated light from a light emitting diode, LED, can be valid as long as we generate photons with a separation larger than the coherence time of the source. 

There are many different technologies that can generate single photons and detect them \cite{BC09,EFM11}. Photomultiplier tubes (PMTs), single photon avalanche photodiodes (SPADs) operating in the Geiger mode or superconducting nanowire detectors are some of the most popular detectors, but there is a growing number of alternatives \cite{Had09}. For instance, there have been important advances in silicon detectors \cite{GGR07} that open the door to integration in electronic circuits and in superconducting nanowire single-photon detectors that extend the high-efficiency detection wavelengths to the near infrared \cite{MVS13}.

Traditionally, while binary decisions between no photons and one or more photons are relatively easy to take, single photon detectors have limited photon counting capabilities. There are new improved detectors, but their cost is still high and most applications use a binary approach to photon detection. Another limitation to most single photon detectors is the time they need to recover after a detection, known as dead time. We will later see how these limitations affect our quantum random number generators. 

\subsection{Branching path generators}
\label{branching}
\begin{figure}[ht!]
\includegraphics{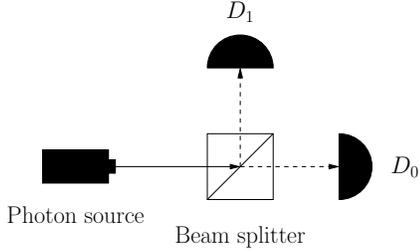}
\caption{\label{BSRNG}A weak light source sends a state with one photon to a balanced beam splitter. The path the photon takes at the output is random and there will be a detection with the same probability at each detector. We can consider that a click on detector $D_0$ is recorded as a 0 bit and a detection in $D_1$ is a 1.}
\end{figure}

OQRNGs take advantage of the random nature of quantum measurement. In a large number of quantum random number generators this measurement is taken over photons in a superposition of two or more paths. For instance, if we define a state $\ket{1}_1\ket{0}_2$ which represents one photon in the first of two possible paths and a state $\ket{0}_1\ket{1}_2$ with the photon in the second path, we can prepare a superposition 
\begin{equation}
\frac{\ket{1}_1\ket{0}_2+\ket{0}_1\ket{1}_2}{\sqrt{2}}.
\end{equation}
Measuring that state with a detector at the end of each path will result in a click in just one of the detectors with a probability one half for each path. There are many quantum optics experiments that generate similar states in Mach-Zehnder interferometers and related optical setups. Figure \ref{BSRNG} shows the archetypal QRNG that uses quantum measurement with detectors in different positions as proposed for the choice of basis in QKD\footnote{In the most popular quantum key distribution protocols, like BB84 \cite{BB84}, E91 \cite{Eke91} or SARG04 \cite{SAR04}, the receiver must choose its measurement basis at random. We can imagine a switch connected to an RNG that directs the incoming photons to one of two alternative measurement setups depending on the result. In practice, the implementation might be different.} \cite{ROT94}. 
In this configuration, we have a balanced beam splitter with equal transmissivity and reflectivity $T=R=\frac{1}{2}$ so that classical light entering any of the two input ports would be divided into two streams of the same optical power, half going through and half reflecting. If we have a single photon in one input and the vacuum in the second, we cannot divide the power and we have the desired path superposition. Conceptually, the simplest way to produce random numbers from this path division is placing two detectors $D_0$ and $D_1$, one for each output, and generate a bit every time we detect a photon. Clicks in $D_0$ would produce a 0 bit and clicks in $D_1$ would produce a 1. Optical QRNGs using spatial superpositions usually apply variations on this basic scheme. In fact, in the original QKD application \cite{ROT94} the random number generator was not fully implemented as a separate device controlling the measurement basis in the receiver. Instead, they used a passive implementation where the beam splitter took the input state and sent it with equal probability to one of two measurement setups, one for each possible basis. A complete implementation with a beam splitter and two photomultiplier tubes as detectors was first deployed as a subsystem in the experimental implementation of a Bell test \cite{WJS98} and later developed as a standalone device \cite{JAW00} with some modifications. The most important difference is the way the random sequence is created, with a random digital signal as an intermediate step. In the modified model, detections in $D_1$ take a digital signal to a high level and detections in $D_0$ to a low level. The result is a random signal with changes in a time scale of the order of the inverse of the mean photon detection rate. If we sample this signal with a clock with a frequency sufficiently below the photon detection rate, assigning a binary 0 when the state is low and a 1 for high state, we obtain a constant stream of random bits. The same procedure was tested with polarized photons in a linear $45^\circ$ state and a polarizing beam splitter with essentially the same results. In \cite{WLL06} there is an alternative take on polarization to path conversion with a weak laser source with linear polarization attenuated to the single photon level and a Fresnel prism that separates the positive and negative circular polarization components and directs them to two avalanche photodiodes. This kind of polarization generator can be modified to provide adjustable probabilities for each bit value if we include an electronically controlled polarizer at the source, like in the fiber-based QRNG of \cite{XHL15} or the decision making system in \cite{NBD15}, which adapts the probability to previous results.

Other generators are implemented in optical fiber systems where a weak light pulse is directed to a balanced fiber coupler connected to two detectors. Two examples are the generators in \cite{SHH01,SHH03}, which use a pulsed laser source that produces, after a tunable attenuation circuit, a coherent state with an amplitude greater than 1 that maximizes the random bit generation rate\footnote{Ideally, we should choose the amplitude of the coherent state $\alpha$ so as to maximize the probability of only one detector clicking either due to one or more photons. For the coherent state at the input of the beam splitter, this amplitude should be $\alpha=\sqrt{2\ln 2}$, but the final configuration uses a  higher level due to additional losses.}.

There are also implementations based on polarization inside optical fiber, with sources that are either single photon states or polarization entangled states
\begin{equation}
\frac{\ket{H}_1\ket{V}_2-\ket{V}_1\ket{H}_2}{\sqrt{2}}
\end{equation}
that are a superposition of horizontally polarized photon states $\ket{H}$ and vertically polarized photons $\ket{V}$ \cite{FB06,FMS06,FSS07,BSS09}. The generators with entangled states produce the photons in nonlinear crystals and use coincidence detectors. One of the photons can be used as a herald or we can watch for anticorrelated polarization measurements in the different paths.  

QRNGs with optical path branching can show a few problems. All types of photodetectors have some kind of dead time after a click. This can generate anticorrelation of neighbouring bits. A detection at some time makes it less likely to find a photon immediately after due to the ``blunted'' sensitivity of the detector before full recovery. Also, for real detectors and beam splitters we will find slightly different detection efficiencies and coupling ratios that can introduce some bias. There are a few other concerns: afterpulsing can create correlated bits, pulses with multiple photons can produce simultaneous detections and the presence of dark counts means there will be occasional clicks when there are no photons. In practice, these effects, particularly dead time, limit the maximum generation rate to a few Mbps, which could be improved with detectors with a smaller recovery time. 

There are many ways to counteract these problems. For instance, the generator in \cite{JAW00} includes a setup phase in which the tube voltage and the detection threshold of the photodetectors can be adjusted to compensate detection efficiency and path coupling differences. Another popular method is applying an unbiasing algorithm that distils a random sequence at the cost of losing some bits. We discuss unbiasing in more detail in Section \ref{postprocessing}. 

If we convert path superpositions into time superpositions we can use one detector instead of two, or more, detectors, and avoid problems caused by having different detection efficiencies and dark count numbers. That is the approach in \cite{SGG00} where weak light from a timed pulsed laser inside an optical fiber is coupled into two fibers of different length connected to the same detector. The additional delay in one path permits to distinguish the route of the photon. The whole attenuation is designed to make each path equally likely. 

The random bit generation rates can improve if the generator measures more than two possible paths. Each measurement then gives more than one random bit. W-states of the form  
\begin{equation}
\ket{W_n}=\frac{\ket{10\ldots 00}+\ket{01\ldots 00}+\ldots+\ket{00\ldots 01} }{\sqrt{n}}
\end{equation}
can be created by branching the photon path many times and give the desired statistics. This approach takes more complex devices, but integrated optical circuits inside silicon chips can offer an economical and scalable alternative. Integrated circuits show less variability and the optical couplers that replace the beam splitters show smaller deviations from a perfectly balanced device. There have been experimental demonstrations of integrated generators with 8 outputs that can produce 3 bits per each measurement, with potential for straightforward extension to 16 outputs \cite{GHP14}. 

Another important point is the choice of photon sources. In many of the reviewed generators, the photons come from LEDs. In order to guarantee independent photons, the rate is limited to be much smaller than the coherence time, which is usually not a problem as the limiting factor tends to the be the dead time of the detectors. A common alternative is using weak laser light. However, it can be interesting to study other photon sources. The effect of a beam splitter on the different quantum states of light is well known \cite{OHM87,PSM87,FL87} and the resulting counting statistics can be used in a variety of generation schemes. There are results that suggest that true single photon sources, which show photon antibunching, can increase the rate of random bits when compared to coherent light from lasers. Brighter sources have a faster photon rate and, in those conditions and once all the effects are considered, single photon sources offer the best overall random bit rates \cite{OG15}.

Finally, there are QRNGs that give up beam splitters altogether. These generators use the natural spatial uncertainty in the generation process. For instance, the commercial Quantis RNG has two integrated detectors placed in positions where the spatial profile of a light source has an equal amplitude \cite{RG09}.

A detector array allows a higher generation rate with more of one bit per detection. In that case, there must be some compensation for the non-uniform spatial profile of most photon sources. An early incarnation of this concept was the optical random number generator of \cite{MM91} that used photon counting detectors with levels around the thousands of photons and needed involved calibration procedures. More recent OQRNG use detectors with single photon precision. One of such generators uses a micro channel plate detector and a wedge and strip anode to assign two coordinates to the place where a photon from an attenuated LED reaches a photocathode \cite{YZL14}. Then, the random bit sequence is extracted from the position using Huffman coding to compensate for non-uniformities. 

Other implementations use an integrated array of single-photon avalanche photodiodes, SPADs, combined with postprocessing \cite{SBC13,BSM13,BSM14}. A weak light source produces clicks in random positions of the array. We can assign a 1 to the pixels that find a photon and a 0 to the pixels that do not click. Even if the distribution of bits in the discrete 2D grid of the detectors is not uniform, we can extract a random sequence if we compare two neighbouring pixels, which should have almost the same probability of detecting a photon, and then take the logical XOR of the bits as their output. Alternatively, we can use the whole string from the array as the input of a randomness extraction algorithm. In these generators, apart from the usual dead time, afterpulsing and dark counts concerns, we have to contemplate the possibility of crosstalk between detectors. However, the effects of crosstalk can be minimized with a proper design. 

\subsection{Time of arrival generators}
\label{timearrival}
There are also multiple ways to use the randomness in photon detection times to generate random bits. The OQRNGs in this and the following section are usually based on the same principles as the QRNGs that detect radioactive decay we discussed in Section \ref{Radioactive}. In fact, one of the earliest proposals for this kind of quantum random number generator was a random pulser that tried to simulate the arrival of radioactive counts in order to calibrate nuclear instruments \cite{TN83}. Some methods are essentially the same than their Geiger counter predecessors but replace radioactive materials with light sources, which can achieve much higher bit rates. Photon production is faster and less problematic and the maximum bit rate is now limited by the capabilities of the detectors instead of the generation speed. 

The basic QRNG using time has a weak source of photons, a detector and timing circuitry that registers either the precise time of each detection or the number of clicks in a fixed period of time. In short time periods with one or only a few photons on average, both the photons coming from LED incoherent light and from the coherent states from a laser arrive at the detector following an exponentially distributed time $\lambda e^{-\lambda t}$ for an average number of photons per second $\lambda$. The time between two photodetections is the difference of two exponential random variables, which is also exponential. In that case, we can compare the time differences between the arrival of consecutive pulses and compare two time differences $t_1$ and $t_2$. We can assign a $1$ if $t_2>t_1$ and a $0$ if $t_1>t_2$. This gives a uniform random bit.

In time of arrival generation, precise time tagging becomes important. Measurement will always have a limited precision and the effects of digitizing the time intervals can be noticeable. Instead of having real times $t_1$ and $t_2$, we have integers with the number of the counted clock 
periods $n_1$ and $n_2$. For instance, the possibility $t_1=t_2$, with a negligible probability for an ideal continuous time measurement, must be taken into account. Now we can find two consecutive measures for which we read the same time, $n_1=n_2$. In our basic scheme that generates a $0$ or a $1$ depending on whether the second interval is shorter than the first one or not, the output is not defined and we must discard these results. Considering the equality as a valid result would require a different analysis of the probabilities of each outcome and how we assign them to a binary bit. 

\begin{figure}[ht!]
\includegraphics{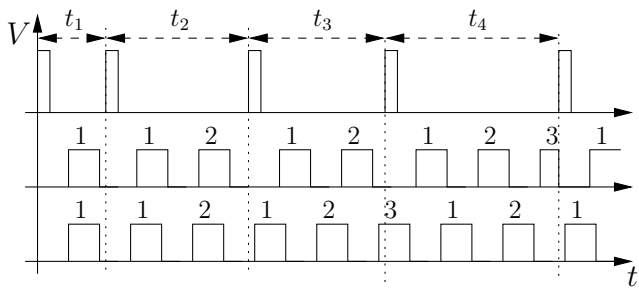}
\caption{\label{digitizing} Generation scheme where the arrival of the rising edge of a detection pulse (up) starts a count of the rising edges of a clock. The clock can be independent from the pulses (bottom) or be reset with every incoming pulse (middle). In the example, $t_2>t_1$ and $t_4>t_3$ and the output should be $11$. Using a resettable clock we find discrete times $n_1=1$, $n_2=2$, $n_3=2$ and $n_4=3$ that produce the sequence $11$ ($n_2>n_1$ and $n_4>n_3$), while for a fixed clock we read $n_1=1$, $n_2=2$, $n_3=3$ and $n_4=2$ and the output is $10$.}
\end{figure}

Figure \ref{digitizing} shows two potential approaches to timing with resettable and non-resettable clocks.

The fine details are explained at length in \cite{SR07}, which gives one of the first optical quantum random number generators that uses time detection. This generator takes the photons from an LED arriving at a PMT and compares the times of arrivals, in a scheme similar to the method that compares the time of arrival of two particles at a Geiger counter shown in Figure \ref{TimeDiff}. As expected, a fast clock with many ticks per click gives better results as we have a higher resolution. A second conclusion is that using a resettable clock eliminates many biases coming from imprecise time measurement. 

A similar generator where the source of the photons, an LED, and the detector, an SPAD, are integrated side by side in the same chip is described in \cite{KEH15}.

The random time of arrival can also be used as a signal that chooses a time bin from a clock, following the template of the radioactive decay generators summarized in Figure \ref{CounterFastClock}. The generator of \cite{DYS08} uses a gated APD detector and outputs a $1$ if a photon is found in an even clock cycle and a $0$ if it is found in an odd cycle. The scheme also adds a self-differentiating circuit to avoid biases from the capacitive response of the detector. An interesting variation on the even-odd generation method is given in \cite{MXW05}, where a pulsed laser produces attenuated states with a small probability of having one or more photons in each time bin. The bins are grouped into pairs and output $0$ is assigned to an empty bin with no detection followed by a detection and output $1$ to a detection followed by an empty bin. This is basically equivalent to using the parity of the time bin where a photon is found, but discards occasional consecutive counts and can be extended to different ways of grouping the time bins \cite{YYW10}.

There are many other proposals that try to generate random bits from time measurements. In principle, each time difference $t_i$ is a real number and it would seem we can extract an infinite amount of entropy from two pulses. However, time precision limits how many usable bits we have. If our timing information has $p$ bits of precision, the time bin in which we find a photon is a random variable with $N=2^p$ possible values and we can compute the probability of a photon arrival in each bin. We can then compute the relevant entropy measure (Section \ref{EntEstimation}) for our discrete probability distribution to see how many bits of randomness are available. 

Certain OQRNGs use digitized time differences with $k$ bits and distill the available entropy into a random bit string with a mathematical function. In \cite{WJA09} the photons from a laser diode are detected with an avalanche photodiode and then the least significant bits of the measured time are collected until they reach 432 bits that are then whitened with the SHA-256 algorithm \cite{NIST12}. Similarly, in \cite{WLB11,WLB14} an attenuated LED sends photons to a photomultiplier tube and the bits from the time of arrival are processed with a resilient function \cite{BR60,SMS07} chosen to take the maximum advantage of the available entropy while doing the processing with a function that can be efficiently implemented in hardware. The generator of \cite{KRK15} also tries to optimize extraction from quantized time differences with hardware designed to work with minimal computation that includes a lookup table that implements Elias' deterministic randomness extraction algorithm (see Section \ref{DetExt} and \cite{Eli72}).

All these processing algorithms try to convert most of the randomness available in the exponential distribution into a uniform bit sequence and require additional hardware and processing effort. 

There are also ways to generate photons with a more uniform time of arrival. The counting statistics at a detector are a function of the photon flux variation at the source \cite{KS68}. For a laser diode with a non-uniform current, we have an inhomogeneous Poisson process and the waiting time at the detector can be adjusted. The generator in \cite{WK10} has circuit that reshapes the exponential time of arrival distribution into an almost uniform one. For a variable photon flux $\lambda(t)$, the time of arrival is a distribution
\begin{equation}
\lambda(t)e^{-\int_a^b \lambda(t')dt'}.
\end{equation}
Ideally, we would want a uniform distribution, which can be approximated by driving a laser with a current that repeats periodically a finite approximation to the function
\begin{equation}
\frac{1}{T-t}
\end{equation}
where $T$ is a reset parameter that determines when to restart the pulse cycle at the source. The current goes back to the initial value when $T$ finishes or when a pulse is detected, whichever happens first.

An alternative way to ``flatten'' the exponential distribution is taking short time bins from an external time reference and consider the time of arrival within those bins \cite{NZZ14}. The time when the photon arrives with respect to the origin of a particular bin is a random variable in a short, almost flat, part of the exponential time distribution, which gives a distribution closer to that of a uniform random variable. 

There are also mixed generators that use both time and space uncertainty. For instance, the generator in \cite{LWW13} uses detectors in two paths to start and stop a timer, in a method similar to the intermediate signal generator in \cite{JAW00}, and uses the resulting time to generate random numbers. In order to have a uniform probability, the scheme assigns a binary string to non-uniform ranges of time measurements that have the same probability. The generator in \cite{TWN08} works with the same kind of intermediate signal. It uses polarized photons combined with a fast clock sampling method (Fig. \ref{CounterFastClock}). The value of a counter is measured with the falling edge of a signal with its transitions controlled by two spatially separated detectors, although there seems to be no post-processing to avoid correlation in the most significant bits. The generator in \cite{SB15} combines a branching path configuration at a beam splitter with the time difference method. There is one random bit associated to the detector that finds the photon and a bit associated to the difference between times of arrival at the detectors. The generator combines both bits to provide a random stream without the biases of the two independent generation methods.

\subsection{Photon counting generators}
\label{photoncounting}
Another large group of generators based on time effects use the number of registered detections in a fixed time $T$. For an exponential time random variable, the number of photons that arrive in a fixed time $T$ follows a Poisson distribution. The probability of finding $n$ photons in that interval is
\begin{equation}
\label{Poisson}
P(n)=\frac{(\lambda T)^n}{n!} e^{-\lambda T}.
\end{equation}

For instance, the generator in \cite{FWN10} follows an approach similar to the radioactive decay generator of \cite{Sch70} (see Figure \ref{CounterSlowClock}) and generates bits from the parity of the total counts registered in a fixed period. The source of light is an LED and, as in many other time-based QRNG, the authors turn to PMTs for faster detection. Interestingly, the generator takes advantage of the dead time of the detector. For the parity method, the random variable that describes the true rate of photocounts when the detector has a small dead time gives a smaller final bias when compared to a pure Poisson process. This approach of taking the least significant bit of the photon count is also followed in \cite{LAV14}, where thermal and weak coherent state sources are compared.

Certain generators use an approach similar to the time difference comparisons of the previous section. If the first measurement gives $n_1$ photons and there are $n_2$ photons in the next time bin, we can generate a $1$ when $n_1>n_2$ and a $0$ if $n_1<n_2$ \cite{RWL11}. 

With these methods we are generating just one bit for each measurement. But, depending on $\lambda T$, our measurements can have a higher entropy. There are some ways to take a fuller advantage of the data we already have. 
 
Certain generators assign more than one bit per detection depending on the counted photon number. The possible results are grouped into sets with equal total probability, which usually requires adjusting the mean photon level of the source to make sure all the sets are really equally, or almost equally, likely \cite{JRW11}.

Depending on the exact photon rate $\lambda T$ in the observed period $T$, the second, third or further least significant bits of the number of counted photons might also be uniform. This is taken into account dynamically in the generator of \cite{TVG15} which has an array of integrated CMOS SPAD detectors that receive light from an LED to generate random numbers in parallel in a $32\times 32$ detector matrix. This is the principle behind the commercial generator of Micro Photon Devices \cite{MPD}. In this approach it is important to properly characterize the dead time, as the rate that registers at the detector $\lambda_{det}$ is affected by dead time. The corrected rate
\begin{equation}
\lambda_{det}=\frac{\lambda}{1+\lambda \frac{t_{dead}}{T}}
\end{equation}
helps to adjust the choice of how many bits from the counted number of photons should be used.

There are also generators that use everyday devices. Certain commercial cameras that are not designed for quantum detection can, nevertheless, offer good enough precision for quantum random number generation. There have been demonstrations of random numbers generated on a mobile phone \cite{SMZ14} from the variations in the count statistic of a state with around 410 photons. In that implementation, the results are taken to a randomness extractor to eliminate correlations and noise effects. This approach is related to the shot noise generators of Section \ref{vacuum}. 

Other photon counting methods take bins of length $T$, subdivide them into smaller bins where we are likely to have zero or one photons and then use more involved procedures to convert the non-uniform Poisson statistics of the large bin into a uniform random variable \cite{WWC15,WXZ15,YZH15}.

\subsection{Attenuated pulse generators}
\label{attpuls}
Certain generators are based on a simplified version of the previous methods with more relaxed requirements for the detectors. Most current single photon detectors have a limited photon number resolving capability and have a binary response of click (one or more photons are detected) or no click (no photon has been found). Photon counting methods usually rely on multiple clicks in a long time period that is divided into a concatenation of smaller bins in the time resolution of the detector. These methods assume a weak source that produces zero or one photons in that bin and that there is a small or ideally negligible probability of generating two or more photons in that shorter time period. 

We call an attenuated pulse generator to the OQRNG with a weak source that has the same probability of generating a photon or not. More precisely, we require the complete system to give a detection probability of one half. We can imagine a superposition of the empty and single photon states in the same spatio-temporal mode (the path that goes to a certain detector in a certain time) so that the quantum state of our photon pulse is
\begin{equation}
\frac{\ket{0}_1+\ket{1}_1}{\sqrt{2}}.
\end{equation}
We can associate a $0$ to a no-detection event and a $1$ to a click. The occupied state does not need to have exactly one photon. Any superposition
\begin{equation}
\frac{1}{\sqrt{2}}\ket{0}_1+\sum_{k=1}^{\infty}\alpha_k \ket{k}_1
\end{equation}
with $\sum_{k=1}^{\infty}|\alpha_k|^2=\frac{1}{2}$ is valid. Externally, we just take the $1$s from clicks and do not care if they are triggered by one or more photons. 

Coherent states provide such a superposition and are easy to produce. For a coherent state of amplitude $\alpha$ the probability of finding zero photons is 
\begin{equation}
p(n=0)=e^{-|\alpha|^2}
\end{equation}
and the complementary probability of finding one or more photons (and finding a click in the detector) is
\begin{equation}
p(n\geq 1)=(1-e^{-|\alpha|^2}),
\end{equation}
as can be seen from Eq. (\ref{coherent}). The simplest idea would be to find the $\alpha$ for which $p(n=0)=p(n\geq 1)$, which happens for $\alpha=\sqrt{\ln 2}$. Eq. (\ref{Poisson}) shows any Poissonian source with $\lambda T=\ln 2 \approx 0.693 $ also gives the desired detector probability. 

In practice, the generator works with an effective mean photon number at the detector $\eta \lambda T$, with an efficiency $\eta$ that depends on many factors such as detector efficiency or path losses. The OQRNG can be adjusted by fine tuning of a variable attenuator. This is the model of the generator in \cite{WG09b}. Alternatively, the generator can act on the light source. The OQRNG in \cite{BMF15,BMM15} adjusts the current of an LED in order to have the desired balance. The OQRNG of \cite{SU15} also has an adjustable source to guarantee a 50\% probability of detection, this time inside an on-demand circuit that produces the photon pulses after a trigger signal has arrived.

Even after tuning, there can be residual bias and the system can drift out of the tuned state during operation. The generator in \cite{WG09a} uses von Neumann extraction to address the problem (see Section  \ref{postprocessing}). For two detections with photon numbers $n_1$ and $n_2$, it outputs a $1$ if $n_1>0$ and $n_2=0$ (a click followed by an empty pulse) and a $0$ if $n_1=0$ and $n_2>1$ (no click followed by a detection). The results with two successive empty pulses or two successive clicks are discarded. For a Poissonian source, both bit values are equally likely with a probability $P(n>0)P(n=0)=e^{-\eta\lambda T}(1-e^{-\eta\lambda T})$. The resulting bit rate is at least four times slower, but free from bias. Greater biases result in smaller rates, but the bits still present balanced probabilities.

\subsection{Generators based on quantum vacuum fluctuations} 
\label{vacuum}
Another group of quantum generators exploits the fluctuations in the quantum vacuum state. The vacuum state can be written as a superposition of amplitude quadrature states $\ket{x}$ 
\begin{equation}
\ket{0}=\int_{-\infty}^{\infty}\psi(x)\ket{x}dx,
\end{equation}
where $\psi(x)$ is the ground-state wavefunction. The wavefunction is a Gaussian around $x=0$ so that
\begin{equation}
|\psi(x)|^2=\frac{1}{\sqrt{\pi}}e^{-x^2}. 
\end{equation}

\begin{figure}[ht!]
\includegraphics{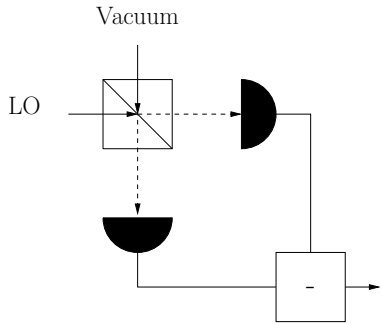}
\caption{\label{homodyne} Homodyne measurement of the vacuum: A laser acting as a local oscillator, LO, is mixed with the vacuum state in a balanced beam splitter. The readings of two detectors at the output of the beam splitter are subtracted and processed to give a current output proportional to the X quadrature of the vacuum field. The proportionality constant is a function of the reference field in the local oscillator.}
\end{figure}

Homodyne measurement \cite{CLG87} offers a simple way to measure the $X$ quadrature. The balanced homodyne detection scheme of Fig. \ref{homodyne} has an output proportional to the quadrature amplitude of the vacuum field and gives an amplified reading of the basic uncertainty in the vacuum state. 

The homodyne detector mixes the vacuum state with a reference laser field from a local oscillator and subtracts the current measurements of two amplitude detectors. The resulting signal can then be processed and digitized to produce the random numbers. Depending on the digitizer that receives the values from the optical detectors, the choice of the local oscillator, the detectors' bandwidth, noise factors and other problems, we might have a different amount of available random bits. With an adequate treatment, the uncertainty in the final measurement can be mostly attributed to the intrinsic quantum fluctuations of the observed vacuum state and not to the shot noise from the local oscillator or other noise sources \cite{YC83}. This random signal can be digitized and sent to a comparator or an entropy extraction circuit to produce random sequences \cite{TV07}. The generator in \cite{STZ10} implements this method by sampling the filtered shot-noise signal periodically and taking the last bit of its digitized amplitude.

We can also take the quadrature measurement, divide the range of possible values of $x$ into boxes from $x_i$ to $x_{i+1}$ and then assign to each box different random bit values. The continuous quadrature value $x$ is in box $i$ with a probability
\begin{equation}
\int_{x_i}^{x_{i+1}}|\psi(x)|^2dx.
\end{equation}
The QRNG of \cite{GWS10} implements this method. It takes 5 bits per measurement (32 bins) and hashes the resulting sequence to remove residual correlations.

QRNGs that measure the vacuum fluctuations can go beyond the Mbps rates of single photon detection methods and reach rates in the Gbps range. They can use fast classical detectors and we can optimize the speed of the electronic part of the generator and concentrate on reducing the technical noise, like the generator of \cite{SAL11}, which discards the least significant bits as a fast method of randomness extraction after noticing that the most significant bits of the digitized homodyne measurement carry most of the quantum noise.

The method can also be used with the squeezed vacuum state. The generator of \cite{ZHZ12} uses second harmonic generation in a parametric oscillator with no input signal to produce a squeezed vacuum state that presents a larger uncertainty in the measured quadrature. In the squeezed vacuum state, the Gaussian wavefunction
\begin{equation}
|\psi(x)|^2=\sqrt{\frac{s}{\pi}}e^{-sx^2}
\end{equation}
is wider by a squeezing parameter $0<s<1$. Homodyne measurement produces a larger range of voltages and makes conversion to digital strings easier. We can define more voltage ranges and reduce the effects of classical noise. With more squeezing (a smaller $s$) the entropy due to quantum noise increases and the bit rate after randomness extraction can be higher. The generation of squeezed vacuum states is described in more detail in Section \ref{OPO} in relation to QRNGs with optical parametric oscillators.

\subsection{Generators based on the phase noise of lasers}
\label{phasenoise}
\begin{figure}[ht!]
\includegraphics{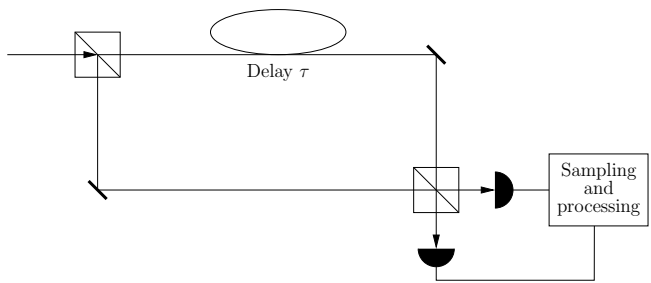}
\caption{\label{MZI} If we divide the light coming from a laser in a beam splitter and make it interfere with a delayed version of itself, the quantum phase noise will produce a random amplitude at the output. Choosing an adequate delay and sampling rate, we can process these amplitudes to generate random numbers.}
\end{figure}

The output of a laser has a random phase of quantum origin that can be used to produce random bits. Inside the cavity of a single-mode semiconductor laser, spontaneous emission causes fluctuations in the output field \cite{Hen82}. This phase noise, also known as phase diffusion, comes from a combination of different quantum effects\footnote{There are many opposing views on the exact role of the vacuum fluctuations and spontaneous emission in laser phase noise and whether spontaneous emission is a direct manifestation of vacuum fluctuations or not \cite{Fai82,Gin83,GSZ88,SS88,HK96}. As far as quantum random number generators are concerned, the exact nature of phase noise is not relevant as long as it is a quantum effect that can produce an observable with a known distribution.}. Direct phase measurement is not technologically feasible for optical signals, but an unbalanced Mach-Zehnder interferometer, MZI, (see Fig. \ref{MZI}) can translate phase differences into amplitude variations. 

In an unbalanced MZI one of the arms introduces a delay $\tau$ with respect to the other arm. Assuming a constant or slowly varying amplitude in each arm, the output has a constant level and a variation proportional to $\cos(\phi(t)-\phi(t+\tau))$ for a random phase difference $\Delta \phi(t)=\phi(t)-\phi(t+\tau)$. The amplitude at the output ports of the interferometer can be measured with high speed standard optical detectors. 

If the introduced delay is far above the coherence time of the laser\footnote{For semiconductor laser with a linewidth $\Delta \nu_{las}$ we can determine a coherence time $\tau_{coh}=\frac{1}{\pi \Delta \nu_{las}}$ \cite{Hen82}.}, $\tau\gg \tau_{coh}$, the phase difference $\Delta \phi(t)$ is a Gaussian random variable of a mean that tends to $0$ \cite{Lax67}. If we sample the amplitude of the detector with a time difference between samples $\Delta t \gg \tau + \tau_{coh}$, the resulting amplitudes are independent \cite{QCL10,GTL10}. These amplitudes are the random variable in many OQRNGs. While the voltages at the detectors carry many classical sources of noise, the quantum phase noise is known to be inversely proportional to the laser output power \cite{Hen82} and, if we operate the laser at a low intensity close the lasing threshold, we can make the quantum uncertainty the dominant noise.  

The generators in \cite{QCL10,GTL10} use the basic configuration in Fig. \ref{MZI} and sample at a fixed period the voltage in one of the detectors. After processing, the voltages $V_k$ measured at times $t_k=t_0+k\Delta t$ are independent Gaussian random variables. 

To generate the random bits, the OQRNG of \cite{GTL10} takes the least significant bit of the voltage measurement or the least significant bit from the difference $V_{k+1}-V_{k}$ between two results if we want to remove biases from the digitization of the voltage amplitudes. 

The generator in \cite{QCL10} adds a phase compensation system in the interferometer to avoid classical phase drift effects that might mask the quantum signal. Its random bits come from comparing each measured voltage with a threshold at the mean voltage value $0$. For the Gaussian voltage signal of interest, we can produce random bits if we choose an output $1$ for $V_k>0$ and a $0$ for $V_k<0$.  

The voltage distribution is Gaussian and we cannot directly use all the digitized bits, which are correlated. However, we can feed them to a randomness extraction algorithm to generate uncorrelated bits. This is the approach of the generators of \cite{LZG10} and \cite{XQM12} which use the same optical delay circuit as the previous implementations and the generator of \cite{NHL15}, which uses a modified interferometer with advanced phase drift correction to achieve rates of tens of Gbps. We can also use Faraday mirrors to correct phase jitter \cite{ZLZ11}. 

For all these generators, we can try to maximize the rate either by increasing the sampling rate or the number of bits we take. However, faster sampling means higher correlations and digitizers have a limited precision. For any given system the randomness rate can be optimized by acting on the sampling rate \cite{ZYM15}. Increasing the sampling rate increases the generated random bit rate until $\Delta t=\tau$. After that point, the bits we read have a higher correlation. The additional samples produce a smaller number of uniform bits and the overall speed decreases. We should choose a delay that maximizes the final bit rate
\begin{equation}
R_s=-\frac{1}{\tau}\log_2 \left [ 2\Phi\left (\frac{\lambda}{\sqrt{\tau}}\right )-1\right ],
\end{equation}
for a parameter $\lambda$ that depends on the laser power, the length of the measured voltage interval and other constants of our system. $\Phi(x)$ is the cumulative distribution function of the standard Gaussian distribution.

An interesting alternative implementation of phase noise quantum random number generators uses pulsed lasers to avoid phase correlations in the optical field. In the generator demonstrated in \cite{JCS11}, a laser is driven by short pulses that take it rapidly from below the threshold to lasing levels. The time the laser is below the threshold, any previous coherence is attenuated and amplified spontaneous emission introduces a new random field. When the laser is suddenly taken above threshold, it amplifies the cavity field to a classical level. After the short amplification stage, the resulting field has a known amplitude due to gain saturation, but the phase is random. 

\begin{figure}[ht!]
\includegraphics{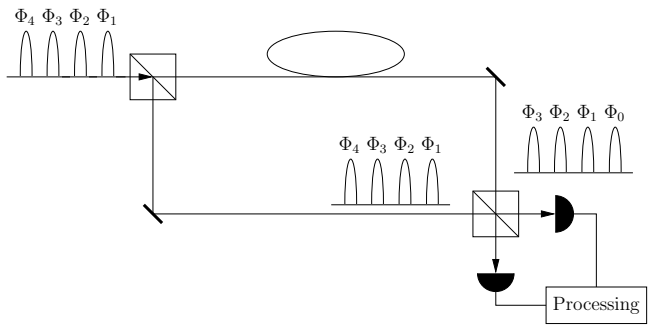}
\caption{\label{MZIpulses} Using a pulsed laser we can generate individual pulses with a random phase due to quantum phase noise. If we introduce a delay in one arm of an interferometer, we have the interference of two pulses with independent phases and the output will have a random amplitude.}
\end{figure}

The resulting output has a series of pulses with a random phase. The phase is converted into amplitude with the usual unbalanced Mach Zehnder interferometer, this time with a delay $\tau$ that matches the repetition rate at the laser so that two consecutive pulses interfere at the output beam splitter (Fig. \ref{MZIpulses}). 

The phase of each pulse $\phi_i$ is uniformly random in $[-\pi,\pi)$ and so is the phase different between neighbouring pulses. The interferometer converts the phase into an amplitude variation that, after detection and filtering, provides energy measurements that are almost uniformly distributed in a restricted range. 

The same configuration with a pulsed laser has been refined later adding passive phase compensation to reduce classical phase drift \cite{TJS13} and tuning the system to achieve a faster rate up to 43 Gbps \cite{AAJ14}.

Quantum noise inside semiconductor lasers plays also a role in classical random number generators based on chaos. There have appeared many random number generators that have one or more semiconductor lasers with optical feedback. The lasers produce a chaotic signal with pulses of a random amplitude and time position \cite{UAI08,RAR09,KAR10,HYM10}. Quantum noise in the laser is the origin of a random variation in the cavity that is then amplified in a chaotic process. While these generators have some entropy due to quantum effects, most of the unpredictability of the final sequence rests on chaotic evolution, which is deterministic. In a sense, they work as physical pseudorandom number generators that take a random quantum seed and expand these small fluctuations at the quantum level into a fast changing physical process to achieve generation rates up to hundreds of Gbps. 

\subsection{Generators based on amplified spontaneous emission}
\label{ASE}
Fiber communication systems owe their fast long range data rates to optical amplification. There are different technologies for optical amplification, like erbium-doped fiber amplifiers, EDFAs, and semiconductor optical amplifiers, SOAs, both popular alternatives in optical communication systems. These optical amplifiers work on variations of the same principle: the light is directed into a medium with population inversion so that the photons in the signal stimulate the coherent emission of new photons that increase the signal's power. However, any excited medium capable of stimulated emission also shows spontaneous emission. That means there appear spontaneously emitted photons inside the gain medium that are amplified by stimulated emission just like the signal is. The random quantum phenomenon of spontaneous emission is thus amplified to a measurable signal with a random amplitude.  

This noise, known as amplified spontaneous emission, ASE, noise is a major limitation to optical gain in communication systems. Larger gains introduce larger noise powers and there is a maximum amplification that can be obtained without degrading the signal-to-noise ratio. Amplified spontaneous emission, either alone or in its beats with the signal or itself, is a strong source of noise that dominates over thermal noise in the detector or the optical shot noise. ASE noise is a first rate challenge in optical communication systems, but can be turned into a good source of entropy in quantum random number generators. Amplified spontaneous emission gives a readily available strong signal with a quantum origin that can be measured with existing optical equipment at fast rates. Sampling random amplitudes of the ASE field in different frequency bands gives statistically independent random variables, even at high sampling rates. The rate of change is usually much faster than the detection mechanism and the speed of the detectors is the limiting factor to the rate in most QRNGs that sample ASE noise. These devices can achieve generation rates of Gbps.

The first proposed quantum random number generators using amplified spontaneous emission work with commercial equipment from optical fiber communications. The generator of \cite{WSL10} uses as a source of random light a pumped erbium/ytterbium co-doped fiber with no input that generates photons by spontaneous emission and amplifies them on their way to a processing stage with a bandpass filter and a second low noise amplifier. The filter limits the signal in the detector to help it work correctly. The signal is then split into its two polarization components, which are independent, and sent to two square-law photodetectors. The resulting voltage signal is mostly what is known as ASE-ASE beat noise, a signal of a random amplitude, with some residual noise from other sources. These voltages have a known distribution that depends on the shape of the filter. The difference of the voltages is a random variable of mean 0. The random bits come from comparing the voltages after each detector $v_1(t)$ and $v_2(t)$, generating a 1 when $v_1(t_i)>v_2(t_i)$ at the sampling time $t_i$ and a 0 otherwise. The resulting sequence still has some small correlation between bits. To correct that, the generator outputs the exclusive OR of the raw bit sequence with a delayed version of itself. 

The generator in \cite{MSL15} also uses a filtered ASE source, a back-pumped erbium doped fiber, but, instead of two detectors, it works with direct detection in a single avalanche photodiode. For the chosen filter, the spectral bandwidth of the optical signal is larger than the detector bandwidth by a factor of $m$. In that case, the intensity distribution that gives the probability of finding $n$ photons for a source with an average of $\lambda$ photons in the time of detection (the inverse of the detection bandwidth) is \cite{WHJ98}
\begin{equation}
p_{BE}(n,\lambda,m)=\frac{\Gamma(n+m)}{\Gamma(n+1)\Gamma(m)}\left(1+\frac{1}{\lambda} \right)^{-n}(1+\lambda)^{-m}.
\end{equation}  
For a high enough value of $n$ and $m$, the distribution has a large standard deviation and most of the uncertainty in the measured voltage comes from the ASE noise and not from electrical noise. We can generate random numbers comparing the results to a threshold value that gives equal probabilities for values below and above it (0 for values below the threshold, 1 for values above). The necessary threshold can vary during operation due to power changes in the source or a drift in environmental conditions during the time of generation. The resulting bias can be corrected with a randomness extractor.

Each measurement can give more than one random bit. The quantum random number generator in \cite{APD12} also uses a single detector but extracts the random bits from a statistical analysis of the random distribution of the detected voltage. The device generates amplified spontaneous emission in two different implementations, one with an erbium-doped fiber amplifier and another with a semiconductor optical amplifier. In both cases, the signal is directed to an optical attenuator. The whole unfiltered noise signal reaches the photodetector where the noise beats give a Gaussian voltage distribution that is digitized. Discarding the few first most significant bits gives a good quality random signal.

Another group of QRNGs uses superluminescent LEDs as the light source. Superluminescent light diodes are incoherent semiconductor sources with internal optical gain that offer an alternative broadband source of ASE. Their output shows a flat spectrum in a wide frequency range. The noise in separate parts of the spectrum is independent and can be used to increase the random bit rate. The generator of \cite{LCM11} can generate multiple bit streams using a wavelength multiplexed configuration where the light from the superluminescent diode is divided into many channels with bandpass filters for different frequency bands. Each channel ends in a single detector whose output is compared to a threshold to generate the random bits. Each output is then processed by taking the XOR of the bit sequence with a delayed version of itself. The experiment in the paper is for a two channel system, but the method can be extended to multiple parallel streams. 

The QRNG in \cite{LWL14} also uses a superluminescent diode in a refined version of the comparison method in \cite{WSL10}. The filtered ASE noise from the diode is amplified in an EDFA and taken to a balanced detection scheme where the optical output is split in two parts and sent to two detectors, one of which receives a delayed signal. This self-differencing takes part of the processing to the optical signal and gives a more symmetric voltage distribution for which it is easier to define a threshold at voltage 0.

As in other generators, we can also try and use all the samples of the digitized voltage and then use postprocessing with delayed versions of the signal to remove residual correlations \cite{YU13}. In that case, the final bit rate can be improved by over-sampling. If we use a sampling rate above the spectrum linewidth of the detected noise, which is limited by the detector, the resulting bits are correlated, but adequate postprocessing can restore a good quality sequence \cite{LZL13}.

A curious alternative is the RNG of \cite{WTG10} that uses the spontaneous emission from a regular light emitting diode without amplification. With no amplifier, the random directions of emission of an LED makes detection difficult. In order to collect enough light, the light source and the detector are placed in the focal points of an ellipsoidal cavity so that the emitted light is collected into the photodiode's sensitive area. The amplitude fluctuations due to the randomness in the emission times come from many independent events and tend to a Gaussian distribution. The voltage at the detector is then sampled and the bits from the digitizer are unbiased to give a random bit string at the output.

\subsection{Generators based on Raman scattering}
\label{Raman}
The interaction between photons and the quantum vibrational states of certain materials is also a good source of randomness. Some quantum RNG resort to Raman scattering phenomena to obtain the entropy for random bit generation. 

There are two important Raman scattering effects. The first is spontaneous Raman scattering. In spontaneous Raman scattering, SpRS, a photon is scattered when it interacts with a molecular lattice that absorbs or creates a phonon to produce a new photon of a higher or lower frequency. If the scattered photon has a larger wavelength and the energy difference is converted into a phonon we speak of a Stokes photon and when there is an energy gain and an incoming photon and an existing phonon produce a scattered photon of a smaller wavelength we speak of an anti-Stokes photon. Anti-Stokes transitions usually produce a smaller field, as they need an established phonon population of the right excited levels of the medium, which in thermal equilibrium is smaller than the population of the ground state \cite{Boy08}. 

Another Raman effect is stimulated Raman scattering, SRS. In stimulated Raman scattering a photon of the frequency $\omega_S$ corresponding to the energy difference between a pump photon and the matching phonon in a spontaneous Raman scattering event stimulates the production of a new photon of the same frequency $\omega_S$. This process can be used to obtain optical amplification. If we have a strong optical pump and a signal at the frequency $\omega_S$, the photons of the signal stimulate the emission of new photons that join the signal pulse consuming phonons and the photons from the pump. This mechanism is used in many photonic devices for amplification and wavelength conversion \cite{Isl02,JRD06} as well as in multiple applications in spectroscopy \cite{CDW90}. While SpRS is almost isotropic and happens at many frequencies, the resulting field in stimulated Raman scattering is mostly contained in a narrow spatial direction and consists primarily of Stokes photons \cite{Boy08}.

Some of the QRNGs based on Raman scattering work on principles similar to those of the amplified spontaneous emission noise generators of Section \ref{ASE}, but, instead of employing quantum spontaneous emission events that are amplified through stimulated emission, they have a strong pump with no input signal so that the spontaneous Raman scattering photons that are produced at random from quantum noise are amplified in a stimulated Raman scattering process \cite{PLK79}. The process starts from spontaneous emission to the Stokes field that comes from the fluctuations of the phonon field \cite{RW90}. The spontaneously generated photons induce new Raman scattering processes and the field is amplified to a macroscopic level in what is known as spontaneously initiated stimulated Raman scattering, SISRS. The quantum fluctuations at the initiating process show at the output field as an uncertainty in the optical phase \cite{KSR91,SBW91,BSW93} and amplitude (photon number) \cite{RRM82,WR83}.

\begin{figure}[ht!]
\includegraphics{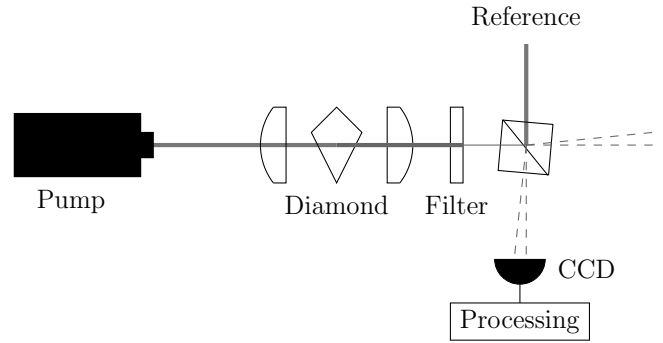}
\caption{\label{diamondpha} Generation of random numbers based on Raman scattering by measuring the phase in the field out of a pumped diamond. In this method, the phase is measured using the interference pattern of the scattered field and a reference. The pattern comes from a tilted beam splitter.}
\end{figure}

The first proposal for random number generation with stimulated Raman scattering \cite{BML11} is based on measurement of the random phase in the field out of an optically pumped diamond (Figure \ref{diamondpha}). Diamonds are a good material for Raman experiments due to their high Raman gain and their transparency at a wide range of wavelengths. A pulsed laser signal is focused into the diamond and produces a Stokes field with a random phase that is uniformly random in the $[0,2\pi)$ range. An optical bandpass filter takes away the pump, which is in a different frequency band than the Stokes field. The random phases are converted into interference patterns at a CCD camera by combining the Stokes field and a reference pulse in a beam splitter. The beam splitter is tilted so that there appear intensity fringes at the detector. The random phase is recovered by fitting the interference pattern to a cosine model and then it is assigned to a bin out of 64 possible phase ranges. The resulting 6 bits are then taken to a bit extraction algorithm to remove any remaining bias. 

The random fluctuations in the amplitude of the field permit a simpler detection scheme without phase to amplitude conversion. Direct detection gives a straightforward amplitude measurement. There is, however, a new problem. Power fluctuations in the pump pulses can mask the quantum effect we want to measure. The generator in \cite{BEN13} monitors the pump power to solve this problem (see Figure \ref{diamondpha}). The basic setup is essentially the same as in the phase Raman random number generator we have just covered. The pump starts an SIRS process in a diamond and the Stokes field is filtered from the pump background. Now we can directly use a detector with the output field. During normal operation, the amplitude fluctuations can reach up to multiple times the mean energy. The exact amplitude distribution has no known analytic expression and depends on the Raman gain, the focusing geometry and the effects of phonon decay, among others \cite{RWM85}. The output field has also small contributions due to pump coupling to more than one spatial mode and other masking effects. The amount of available entropy can be estimated deconvolving the Stokes energy distribution from the detection noise, as measured without a signal. The results show only a small effect of electrical detection in the total noise. In order to extract the entropy, the measured intensity values are corrected with the power values of the monitored reference and the compensated amplitude measurements are binned into intensity ranges that are assigned a bit string. As a last step, the sequence is applied Toeplitz hashing to remove bias and classical noise.

\begin{figure}[ht!]
\includegraphics{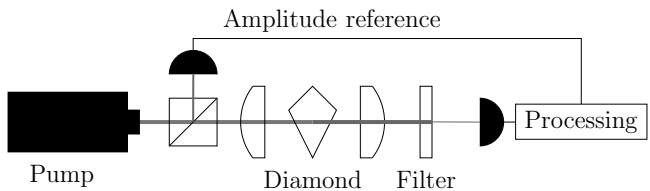}
\caption{\label{diamondamp} We can use the amplitude fluctuations in Raman scattering as a randomness source. In order to correct for the fluctuations of the pump, which do not have a quantum origin, we must include an amplitude correction method.}
\end{figure}

In both cases we have described, Raman interaction has a potential for fast generation rates. The system dephases in times of the order of a few picoseconds, resetting the vacuum phonon state before the new random field is generated. The pulses come with a period much longer than the dephasing time for the phonons in the diamond. In these random number generators, the rate limit comes from the repetition rate of the laser. Stimulated Raman scattering requires large powers in order to produce a strong output signal. In the free-space configuration of the discussed generators the available lasers limited the rate to the range of kbps. These rates can be improved with faster lasers.

An alternative way to measure phase differences with a higher rate is given in \cite{EBM14}, where Raman interaction happens inside a highly nonlinear Potassium Titanyl Phosphate, KTiOPO$_4$, waveguide. Wave\-guides offer tight confinement and the guided pump field has a stronger interaction with the medium that allows us to use power levels in the range of faster repetition lasers, like the titanium:sapphire oscillator with a repetition rate of 80 MHz of this generator. The random numbers come from converting the random phase into an amplitude variation in an interferometer with a delayed arm, like in the schemes for quantum random number generators based on phase noise we discussed in Section \ref{phasenoise}.

The quantum effects in spontaneous Raman scattering, SpRS, can also serve as a randomness source in schemes without amplification at the cost of adding single photon detectors. By improving the detector, we can have a continuous wave laser pump of relatively low power. If we only observe the scattered photons with large frequency shifts, this interaction is mostly between the input photons and the vacuum noise phonon fluctuations instead of interactions with the thermal phonon field. The quantum randomness from phonon vacuum fluctuations is the principle behind the QRNG of \cite{CCY14,CCX15} where a strong pump inside a highly nonlinear As$_2$S$_3$ fiber generates spontaneous Raman photons in different frequency bands. The pump photons interact with phonons of different energies. The scattered photons occupy the spectrum following a known probability distribution with two separate regions. One part of the spectrum is associated to thermal phonons and, in a time $T$, it has an expected scattered photon detection rate \cite{CCX15,KS73,LYA07}
\begin{equation}
R(\nu,T)=C\eta \Delta \nu P L(1+n_{BE}(\nu,T))g(\nu)
\end{equation}
that depends on different experimental parameters like the Raman coupling efficiency $C$, the experimental loss factor $\eta$, the measurement bandwidth $\Delta \nu$, the laser power $P$ or the effective scattering length of the device $L$. Two particularly interesting factors are the gain profile of the medium with frequency $g(\nu)$, which includes both polarizations, and the thermal phonon occupation number 
\begin{equation}
n_{BE}(\nu,T)=\frac{1}{e^{\frac{h\nu}{k_B T}}-1}
\end{equation}
that gives the Bose-Einstein distribution of the population of phonons with energy $h\nu$ for a thermal energy $k_B T$. This distribution is close to the smaller detunings with respect to the pump. The photon distribution in frequency is concentrated a few THz above the pump frequency.

The spectrum has a second peak at higher detunings due to the quantum vacuum fluctuations of the phonon field. In the discussed As$_2$S$_3$ fiber, the distribution peaks around 10.4 THz above the pump \cite{CJC12}. At room temperature, the distributions of quantum and thermal origin are centered around different parts of the spectrum. While both distributions are random, the thermal component shows the same problems as the thermal noise generators we discussed in Section \ref{noise} and we prefer the more stable random distribution from the quantum part of the spectrum. There is still some contribution from thermal scattering events, but this and other biases can be corrected with postprocessing. 

Once we have selected the most adequate frequencies, we can use a coarse wavelength division multiplexer to measure two slices of the spectrum with an equal probability of having a spontaneously scattered photon. The multiplexer converts the spectrum distribution into a spatial separation. The rest of the scheme basically follows the model of the spatial separation generators we discussed in Section \ref{branching}. In the discussed experiments, two detectors $D_0$ and $D_1$ measure the photons in the paths of the two spectrum slices during a time $T$. The output bit is a $0$ if there is a click only in $D_0$ and a $1$ if only $D_1$ clicks. Two simultaneous detections and empty time bins are discarded. The differences in the collection efficiencies of the two detector channels and the non-flat shape of the Raman spectrum introduces biases in the sequence. In order to correct the bias, there is a postprocessing stage that XORs the sequence with a 16-bit delayed version of itself. 

The experiment gave raw generation rates of 1 Mbps, 650 kbps after postprocessing. The ultimate limit for the random bit generation rate depends on the decay time of the Raman response function. Spontaneous Raman scattering photons that are generated with a time separation less than the Raman response time can have frequency correlations. The photon generation rate can be controlled with the power of the pump laser to avoid correlations. In the studied fiber, the medium reacts in less than 100 fs \cite{AKN95}. Generation rates up to 1 GHz would still show a small two photon probability of the order of $5\times 10^{-3}$ in that response interval. 

In the experiment, the generation rate is limited by the detectors. The detector limitations are the same as in the generators based on single photon detection we discussed in Sections \ref{branching} and \ref{timearrival}. Most single photon detectors are limited to a MHz rate, but more advanced detector technologies can bring the rate closer to the Raman physical limit. Additionally, the rates can be improved by dividing the spectrum into more than two channels. A wavelength division multiplexer can take the photons into multiple paths that allow to extract more than one bit per measurement.

\subsection{Generators based on optical parametric oscillators}
\label{OPO}
Binary phase selection in degenerate optical parametric oscillators offers a further way to amplify quantum randomness from the microscopic level to a macroscopic optical field. In an optical parametric oscillator, OPO, the photons that appear from spontaneous parametric down conversion of the light from a pump start an oscillation inside a cavity, even without any input at the resonant lower frequencies \cite{LYS61,HOB67}. The zero-point fluctuations alone can initiate the gain in the cavity. The principle is similar to the amplification of quantum noise inside a laser we have discussed in Section \ref{phasenoise}.

In spontaneous parametric down conversion, the nonlinear response of a medium converts the photons from a pump at a frequency $\omega_p$ into two photons: a signal photon with frequency $\omega_s$ and an idler photon at $\omega_i$ so that $\omega_p=\omega_s+\omega_i$. This phenomenon has applications in entanglement generation and in parametric amplifiers. In a medium with type I degenerate down conversion each photon from the pump produces two photons with the same frequency and polarization. Different pump photons give different polarizations, but all the generated photons have the same frequency. In these conditions, an optical parametric oscillator with no input but the pump amplifies the uncertainty in the vacuum fluctuations and the output is a squeezed vacuum state where the uncertainty at the quantum level can be measured from a macroscopic optical signal \cite{WKH86,WXK87}. 

The cavity of an optical parametric oscillator has losses and there is a gain threshold below which spontaneous parametric down conversion cannot be amplified to the macroscopic level \cite{YY07}. In a continuous wave type I degenerate OPO where both the signal and the idler fields are indistinguishable, the gain mechanism is phase dependent and has a period of $\pi$ for the signal phase \cite{NYD90,MLP12}. For an adequate pumping power, there are only two stable oscillation states where the gain is greater than the oscillator losses. These states show a phase with respect to the pump around $\theta_s=0$ in one state and around $\theta_s=\pi$ in the other.

The optical parametric oscillator quantum random number generators of \cite{MLV11,MLV12} use as their randomness source the phase of the macroscopic field inside the cavity, which is inherited from the vacuum fluctuations. In this process, classical noise effects are negligible and do not change the phase state. In order to convert the phase variations into a binary random number, we can take two independent cavities of the same output power and make their output fields interfere at a beam splitter. If both cavities have a state around the same phase, there will be a constructive interference and the signal will have close to double the original power. If the phase states are around opposite values there is a destructive interference and the output power is close to 0. 

\begin{figure}[ht!]
\includegraphics{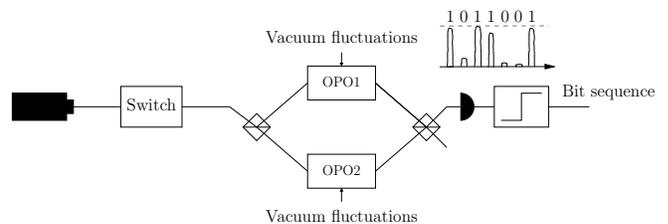}
\caption{\label{twoOPOs} Quantum random number generation with two optical parametric oscillators. A pulsed laser creates an oscillation in each OPO in one out of two possible stable states with a phase centered around $0$ or $\pi$ with respect to the pump. The final stable phase depends on the initial conditions of the quantum fluctuations in the cavity and when the pulses from both OPOs interfere we will have close to totally destructive or close to totally constructive interference. The resulting amplitudes can be easily distinguished and be assigned to the 0 and 1 bit values.}
\end{figure}

For the right cavity parameters, the phase distribution can be quite narrow around the central values $\theta_s=0$ and $\theta_s=\pi$ and the output power of the interferometer has two clearly distinguishable optical power values that can be told apart using a threshold in the middle of the expected detector voltages corresponding to a totally constructive interference and a totally destructive one. The value of the comparison can be used to generate random bits. A low voltage state (destructive interference) can be interpreted as a 0 and a high voltage state (constructive interference) as 1.   

The bit rate depends on the time it takes for the cavity to generate a new random phase. Once a stable state is established inside the cavity, it will feed itself. We need to restart the oscillation to generate a new random value. In the generator of \cite{MLV11,MLV12} the cavity is detuned by blocking and unblocking the pump. 

There is a minimum time before we have a fresh source of randomness. We must first allow the field inside the cavity to decay to the quantum noise level before a new oscillation builds up. Otherwise, when we establish the oscillation, the residual field dominates over quantum fluctuations and the new phase state is correlated to the previous phase value. This is the limiting factor in the speed of OPO-based QRNGs. The exact time for regeneration depends on the cavity and the pump power. If we pump well above threshold, like in the described generators, it can take from 10 to 20 times the $1/e$ decay time of the cavity to go back to the quantum noise level \cite{MLV12}. The intensity decay time can be estimated from the oscillator parameters as
\begin{equation}
\tau_{off}=\frac{T}{2\delta_E-2\delta_E\sqrt{\frac{P_{\text{off}}}{P_{\text{th}}}}}
\end{equation}
for a cavity with an electric field fractional roundtrip loss $\delta_E$, a cavity roundtrip time $T$ and pump powers at the threshold and ``off'' levels of $P_{\text{th}}$ and $P_{\text{off}}$ respectively \cite{MLV12}. 

In the described QRNGs the bit rate is in the order of tens of kbps before serious correlation problems appear. Shorter cavities can have lower build-up times and, when combined with pumps at higher repetition rates, would allow rates in the Gbps range \cite{LPP05}. 

There are also interesting variations of the method with other parametric processes. This generation method is not necessarily restricted to second-order nonlinear materials. Instead, we could use $\chi^{(3)}$ effects in integrated optical parametric oscillators \cite{RDF10,LOV10}. 

Apart from optical parametric oscillators, there are other bistable optical systems where quantum effects can produce jumps between stable states. For instance, the quantum random number generator in \cite{SHA11} uses a semiconductor ring laser that is driven from a monostable to a bistable state. The amplified spontaneous emission noise in the counter-propagating laser modes that appears during switching defines the final stable state from the two possible options and gives a random macroscopic bit that has a quantum origin. 

Competition between optical modes is also the source of randomness in the generator proposal of \cite{SSS13}, in which spontaneously emitted photons in two possible competing modes are amplified in a laser setup so that there is a macroscopic winning mode that amplifies the quantum uncertainty at the single photon level.

\section{Non-optical Quantum Random Number Generators}
\label{Other}
While quantum light offers a simple source of quantum randomness, there have also been proposals for quantum random number generators based on other physical systems. 

For historical reasons, we have already discussed in their own section the quantum random number generators based on the random behaviour of radioactive decay (Section \ref{Radioactive}). They were the first quantum random number generators well before the explosion of quantum information theory and remain in use. While they are based on the detection of particles, they are in many aspects equivalent to the optical schemes based on photon counting, time of arrival and position (in fact, in the case of $\gamma$ radiation we can say we have an optical system, just with photons of a very high frequency).  

A second family of non-optical random number generators with a quantum contribution is the group of electronic RNGs we have covered in Section \ref{noise}. In general, their source of randomness is not so clearly defined as in the rest of quantum random number generators described in this paper, but noise generation with Zener diodes, when implemented properly, can be taken to an almost purely quantum regime \cite{Sti04} and electronic shot noise is the source of randomness in certain commercial quantum random number generator of ComScire \cite{Com}. 

In a reverse-biased Zener diode with a low breakdown voltage, the dominant source for the current that appears is the completely quantum tunnel effect \cite{Pie96}. The p-n junction of the diode presents a potential energy barrier that is thin enough to allow random quantum tunneling of some of the electrons from the valence band of the p-side to the conduction band of the n-side of the junction. This creates a random reverse current that is the basis for many electronic noise physical random number generators. 

Similarly, the tunnel effect at the p-n junctions in MOS transistors creates a leakage current formed by the electrons that tunnel through the insulating layer under the gate. This tunneling introduces a varying current that suffers from shot noise due to the discrete nature of the electrons. These changes can be converted into a variable jitter in ring oscillators and processed to produce random numbers \cite{Com}. The origin of the noise is similar to that of the optical random number generators discussed in Section \ref{vacuum}, but replacing discrete elements of light (photons) with discrete elements of current (electrons). 

The shot noise in p-n junctions of different semiconductor devices is a usual source of randomness in home-made electronic random number generators. An example is the random number generator based on reversed-biased p-n junctions in transistors of \cite{CL15}.

Quantum tunneling is the basic principle behind these and many additional non-optical random number generators. Apart from shot noise in p-n junctions, tunneling explains, among others, cold emission of electrons from metallic surfaces or alpha decay \cite{Raz14}. From that point of view, we can say a QRNG based on radioactive alpha decay is also based on tunneling. Similarly, the random number generator that amplifies the electrons coming from nano-size emitters under an electric field in \cite{VBG11} is a QRNG based on tunneling.

Other quantum random number generators measure the state of atomic quantum systems, like trapped ions. QRNGs based on measurements on trapped ions, while slower than their optical counterparts, have an interesting application to device independent quantum random number generation \cite{PAM10} and other certified generators that are based on experimental tests of quantum mechanics \cite{UZZ13}. Trapped ions systems are more complex to implement than most optical measurement setups, but they offer almost perfect detection efficiencies, which is paramount in certification. Due to the special interest of this generation method, we give a more detailed description in Section \ref{DevInd}. 

There are also more exotic proposals related to the certification of the produced random bits, like generating random numbers with Majorana fermions \cite{DD13}. A Majorana fermion is a particle predicted in 1937 \cite{Maj37} for which there is convincing experimental support \cite{NDL14} and which would have desirable properties against noise and imperfections in certain implementations of quantum information protocols.

Another curious proposal is the QRNG of \cite{KPT08} that measures the quantum fluctuations of the collective spin of an alkali-metal vapor. Spin noise is a random magnetic moment that appears when we have a collection of atoms, even in the absence of an external magnetic field, and is proportional to the number of involved atoms. Spin noise allows to probe the properties of the system efficiently with experiments imitating magnetic resonance methods and its measurement has applications, among others, to spectroscopy in semiconductors \cite{HBD14,KDK07}. 

Spin noise is an Ornstein-Uhlenbeck stochastic process that appears from the quantum uncertainty of the spin degrees of freedom combined with measurement-induced noise coming from atomic collisions. The spin state can be probed optically due to optical selection rules that permit to map the varying spin polarization onto the intensity of a probe light beam. With a proper setup, the fluctuations in the optical power due to spin noise dominate over the electronic noise and the photon shot noise and the optical power gives a precise measurement of the global magnetic field. 

The QRNG in \cite{KPT08} measures the spin noise by analizing the polarization of a probe beam after traversing an alkali-metal vapor under a magnetic field. Spin noise produces a random change in the polarization that can be monitored by measuring the amplitude in the horizontal and vertical component of the light after a polarizing beam splitter. Comparing the level in one branch to a threshold that includes the presence of background noise, we can generate a random binary sequence assigning a 0 or a 1 depending on whether we stay below the threshold or not.

The generation rate reaches the kbps range and is limited by the relaxation time of the system. In this case, it is desirable that the coherence of the system is short-lived so that a new random state can be created as fast as possible. Samples below the relaxation time would be correlated. Nevertheless, there are systems with lower relaxation times, particularly solid state systems like GaAs structures, which could allow dephasing rates in the order of 1 GHz \cite{SZK07,ORH05}.

\section{Random numbers certified by quantum mechanics}
\label{DeviceIndependent}
Cryptographic random number generators face a problem of trust. Users must ultimately trust the algorithm of a pseudorandom number generator or the device that implements a true random number generation method. The alternative, which is devising a random number generation from scratch, is highly undesirable. The cryptographic maxim ``Don't roll your own crypto'' sums up the collected experience of the security community and warns against non-tested systems. Trusted algorithms and devices have resisted years of cryptoanalysis and attempted attacks and public inspection vouches for their robustness. 

Unfortunately, this means that, at some point, users must trust the device or the algorithm they are given. The question, which might seem academic or for the paranoid-minded, is not trivial. The events in the last years have shown RNGs are a tempting target for hidden attacks. For instance, the pseudo-random number generation algorithm DUAL{\_}EC{\_}DBRG, which was proposed as a NIST standard \cite{BK07}, allows backdoors that permit an attacker to recover the whole random sequence with minimal information \cite{SF07,CFN14,Hal14,BLN15}, which has had practical consequences in the Juniper network attack \cite{CVE7755}. At the hardware level, there are demonstrations of how a rogue manufacturer or any attacker with access to the device can insert very hard to detect errors in real world RNGs by introducing dopants in certain parts of the circuit \cite{BRP14}. This is an example of the more general threat of hardware trojans, which are different kinds of malicious modifications that are inserted at the hardware level \cite{TK10}. 

For physical random number generators there is also the possibility of spontaneous failure. If a component from the device stops working or degrades, the quality of the output bits might suffer. Subtle hardware failures can be hard to notice, especially if the device still produces an output. For that reason, security recommendations like the AIS 31 standard of the German \emph{Bundesamt f\"ur Sicherheit in der Informationstechnik} \cite{KS11} or the draft of NIST SP 800-90B \cite{TBK16} ask for some kind of self testing inside true random number generators. A subsystem should monitor the state of the device at all times \cite{BL05,Fis12}.

In this Section, we review three quantum-inspired ways of working with untrusted devices. The first method is using some properties associated to quantum phenomena to observe the quality of the produced bits. The second subsection gathers the proposals collectively known as device independent quantum random number generators, which are based on the clever realization that there are quantum correlations that guarantee certain statistical independence unless some trusted physical principles, like causality, are wrong. The third part describes quantumness certification methods that are inspired by device independent generators, but use less stringent experimental tests of different aspects of the quantum theory and provide a limited certification under more relaxed security assumptions.

\subsection{Self-testing in quantum random number generators}
Most quantum random number generators do not fully characterize their source of randomness. For instance, while a photon at a beam splitter (Figure \ref{BSRNG}) should produce perfectly random bits, there can be problems with detector efficiency, unbalances in the splitting process, imperfections in the source and many unsuspected sources of correlation. For that reason, there have appeared different methods to check the quality of the random numbers produced in physical random number generators. This is not exclusive to quantum random number generators. In classical physical random number generators there are different ways to check the output to detect failures, like including hardware versions of the NIST and Diehard randomness tests we describe in Section \ref{tests} \cite{SAB13,YRM15a,HCS10,SSR09a,VLS11,SSR09b,VLJ10}. Here, we discuss only the self-testing approaches that are directly related to the quantum properties of the random number generator. 

There are also self-testing methods that can work both with classical noise and quantum sources of entropy. The self-testing circuit described in \cite{SIT10} compares the time of arrival of random pulses coming either from thermal noise or from the detection of radioactive decay with a Geiger counter (Section \ref{Radioactive}) and tests the resulting distribution against the expected Poisson time of arrival. Only the random numbers passing the tests are put forward to the output, filtering out obvious failures. 

While there is still a risk from a malicious attacker that modifies the output to produce predictable sequences that will pass the tests, these self-checking systems can detect spontaneous failures and less sophisticated attacks and they are a good addition to security. Tests can serve as a canary to detect operation errors and alert that something is wrong. 

Testing must be done with due care. Accurate entropy estimation is a hard problem and a system that evaluates the available entropy with a poor implementation can be vulnerable to attacks \cite{DPR13}.

The first mention to self-testing in a quantum setting was presented in the optical QRNG of \cite{FSS07} that is designed to work with either a single photon in a polarization superposition
\begin{equation}
\ket{\psi}=\frac{\ket{H}+\ket{V}}{\sqrt{2}}
\end{equation}
or with an entangled state
\begin{equation}
\label{entangledpol}
\ket{\psi}=\frac{\ket{H}_1\ket{V}_2+\ket{V}_1\ket{H}_2}{\sqrt{2}}.
\end{equation}
The quantum random number generator works on the principles of path branching discussed in Section \ref{branching}. 

The device includes a testing phase in which it performs full tomography of the input state \cite{JKM01} from a set of measurements in order to determine the $2\times 2$ matrix that describes the photonic two level system for a single photon or the effective two-dimensional Hilbert space of interest in the case of the photon pair. From the measurement results, the generator estimates the minimum possible min-entropy $\widetilde{H}_{\infty}(\hat{\rho})$ for the joint state of the user and an eavesdropper, $\hat{\rho}$, for the worst case over all the possible decompositions. Then, the raw bits are fed to a randomness extractor \cite{BST03} that, for the estimated bound on the available entropy, produces a shorter unbiased random string.

This method offers protection against an adversary that can control the quantum state from which we obtain the entropy as long as we can take repeated measurements on the same state. In order to perform state tomography correctly, we need to assume the measured state is the preserved throughout the process. This can be interesting when the attacker can only alter the photon source or when there is a physical problem with the generator. While this kind of self-testing offers a limited protection against advanced attackers, it is an effective way to detect accidental errors in the device.

Tomography offers a reasonable entropy estimation in models where we assume honest errors in implementation or failures during operation instead of a collection of components from untrusted colluding manufactures. Such a model is put forward in the self-testing QRNG of \cite{LBL15} where randomness from a quantum origin is separated from technical noise using the dimension witness of \cite{BQB14} defined as
\begin{equation}
W= \left|
\begin{array}{cc} 
p\left(1\!\mid 0,0\right)-p\left(1\!\mid 1,0\right) & p\left(1\!\mid 2,0\right)-p\left(1\!\mid 3,0\right)  \\ 
p\left(1\!\mid 0,1\right)-p\left(1\!\mid 1,1\right) & p\left(1\!\mid 2,1\right)-p\left(1\!\mid 3,1\right)  
\end{array}
\right|
\end{equation}
where $p\left(b\mid x,y\right)$ gives the conditional probability of finding an outcome $b$ (from $\pm 1$) for a state prepared in one out of four $x=0,1,2,3$ possibilities in a measurement setting $y$ that can be $0$ or $1$. In the discussed generator, the four states correspond to the circular right and left polarizations or the diagonal and antidiagonal polarizations of the second photon from an entangled pair, which is measured in the diagonal or the circular polarization basis. The first photon acts as a herald. 

$W$ gives an idea of ``how quantum'' is the combination of preparation and measurement. Any $W>0$ shows that some measurements are incompatible and there is some quantum randomness that allows to give a bound on the guessing probability. The result can be used to decide the level of compression in a randomness extractor. For smaller values of $W$ (a more classical behaviour) the raw input bits produce a smaller number of clean random bits. The experimental test of this method in \cite{LBL15} gave a final bit rate around tens of bits per second and showed a correct response to environmental changes, like the alignment problems resulting from turning off the air conditioning in the lab.

A similar approach to self-testing with a Faraday-Michelson quantum key distribution system \cite{MZH05} is given in \cite{SLY15}. 

An alternative  is to take advantage of the uncertainty principle to ensure any adversary has a limited amount of information. As in the previous methods, our goal is not only to generate random bits, but to be sure they are private (no external attacker can learn our sequence). For instance, if we measure the polarization of the first photon in the entangled state of Eq. (\ref{entangledpol}) in the horizontal-vertical basis, we would get perfectly random numbers, but an adversary that captures the second half would know the exact sequence we obtain by taking the same measurement. This can be acceptable in applications like simulation, but in cryptography we need to avoid any information leakage. The certification method in \cite{VMT14} is designed to ensure privacy without full tomography by switching between two mutually unbiased bases \cite{BBR02,DEB10}. Instead of a full tomographic measurement, two bases are enough. The conditional min-entropy with respect to an eavesdropper (Section \ref{EntEstimation}) gives a bound to the amount of randomness we can safely extract from a measurement \cite{KRS09,DPV12}. The uncertainty principle guarantees there is a limited correlation with the environment for any possible input state (we can prove a bound on the conditional min-entropy from our measurement results). This implementation requires a small random seed to choose between the bases. The original randomness in the seed is expanded after the measurements into a reliable private bit string. The seed needs to be uniform and cannot be taken from the same weak randomness source as the rest of the bits (see Section \ref{postprocessing} for a more detailed description of randomness extraction and the role of uniform seeds). The method was demonstrated with entangled photon pairs generated from parametric down-conversion and measurement in the diagonal/antidiagonal and the horizontal/vertical polarization bases. 

We can also follow the methods of precision measurement \cite{MBN13,BNW14} and propose a complete model of the generator where all the sources of uncertainty are rigorously characterized and all the experimental imperfections are taken into account in the most conservative way. The experimental standards followed in precision measurement have been put to test in atomic clocks with impressive results and can be adapted to quantum random number generation. This characterization based on metrology has been followed by \cite{MAA15} to vouch for the randomness in a phase noise QRNG. The chosen device, described in \cite{AAJ14}, is based in the random phase in a laser, as explained in Section \ref{phasenoise}. A physical model can give a strict bound for the average min-entropy, which is used to choose a randomness extractor. The method works with theoretical considerations alone, but also gives room to introduce constraints based on auxiliary measurements or on the data that has been generated. This kind of estimation has also be done in \cite{HAL15} for the initial configuration of the QRNG based on the measurement of vacuum fluctuations of \cite{SAL11} (see Section \ref{vacuum}).

\subsection{Device independent quantum random number generators}
\label{DevInd}
A second approach to certifying random numbers is ignoring the details inside the quantum random number generator and judge the results based only on the output. In particular, we want to prove that the output must be random or otherwise some physical law must be broken. This is the basic model behind device independent quantum information processing, which started in the context of quantum key distribution with \cite{MY98} and \cite{BHK05} with multiple further developments \cite{Col07,CK11,MMM06,ABG07}.

In the case of random number generation, it tries to address the worst imaginable case where an adversary has generated genuinely random numbers, for instance with a quantum random number generator, and then has hidden them inside a manipulated device. If we check the output of that device, the sequence will pass all randomness tests and we will trust the results. This problem is difficult to avoid, but has a quantum solution.

Device independent quantum random number generators solve the problem of trusting the device with schemes based on Bell tests. The ideas of Bell violation stem from the discussion of an apparent discordance of quantum theory and relativity known as the Einstein-Podolsky-Rosen paradox \cite{EPR35}. In an entangled state, measurement of one of the particles immediately sets the state of the other particle. This seems to contradict the no-signalling principle than forbids faster than light communication. John Bell showed that the contradiction could be settled experimentally \cite{Bel64}. The statistics of measurement on space-like separated entangled particles would be different in a realistic local world with no interaction faster than light and in a world where the laws of quantum mechanics hold. Both alternatives are incompatible. Aspect's experiment \cite{AGR82} showed support for the quantum description. There are, however, experimental loopholes that could still allow a hidden variable theory that is local or realistic. A series of ever more sophisticated experiments is closing alternative explanations and confirm the predictions of quantum theory \cite{GVH15,SMC15,HBD15}. A detailed description of Bell inequalities and nonlocality can be found in \cite{BCP14}.

In the experimental QRNG of \cite{PAM10} the chosen version of the Bell's inequalities is the Clauser-Horne-Shimony-Holt, CHSH, formulation \cite{CHS69}, which is particularly elegant, simple and intuitive. We study the correlations in measurements from two devices and define two variables $x$ and $y$, one for each device. The variables can take two values, 0 and 1, that correspond to a choice between two binary measurements. Both measurement devices are identical. The measurement in the $x$ configuration gives a binary output $a$ and the measurement defined by $y$ gives an outcome $b$. We are interested in the correlation function
\begin{equation}
I=\sum_{x,y}(-1)^{xy}[P(a=b\mid xy)-P(a\neq b\mid xy))]
\end{equation}
where $P(a=b\mid xy)$ and $P(a=b\mid xy)$ are the probabilities that $a=b$ or $a\neq b$ when the settings are $x$ and $y$. For a realistic local theory we should always find $I\leq 2$. Any value above $2$ indicates non-locality.

The function $I$ can be experimentally approximated by estimating the probabilities after taking a series of measurements. As long as the systems are separated and do not interact, if the laws of quantum mechanics hold and the inputs $x_i$ and $y_i$ at any stage $i$ are generated by independent random processes, the estimation of $I$, $\tilde{I}$, gives, after some work, a lower bound to the min-entropy of the outputs. The original derivation of the bound on min-entropy in \cite{PAM10} had a technical error, but in \cite{PM13} and \cite{FGS13} there are restored correct proofs of the main results, as well as demonstrations of some additional properties of the protocol, like its composability\footnote{In cryptography, proofs of security are limited to the particular conditions of the protocol and might fail when the results are put forward to a second cryptographic protocol. Putting together the information leaked from the first and the second protocol can compromise the data in a way neither protocol alone does. We say a protocol is composable if we can prove its output can be safely used as the input of another protocol, maybe under some restrictions. A composable protocol can be used as a part of a larger system and is still secure \cite{Can01,BCN04}.} and its fitness to generate random bits for their use in cryptography.

If the system admits a classical description, $\tilde{I}\leq 2$, the bound is zero and the system could be deterministic. If the measurements are done on states showing some entanglement the produced random bits are guaranteed to have some randomness. The resulting bit sequence is not necessarily uniformly random, but the bound in its min-entropy means it can be converted into a random uniform string with an appropriate randomness extractor (see Section \ref{postprocessing}). 

For quantum devices with spacelike separated parts with access to independent random sources, there are no additional constraints on the devices or the input states as long as $\tilde{I}>2$. The only additional requisite is that the chosen measurement settings $x_i$ and $y_i$ at each stage of the protocol have some randomness (are not perfectly predictable). In that respect, the described generator is a randomness expansion scheme, much similarly to what happens in Ekert's proposal for quantum key distribution \cite{Eke91,VV14}. Starting from a random seed, the protocol gives a longer output random string whose randomness is certified by quantum mechanics. The protocol in \cite{PAM10} is quadratic: in order to produce $n$ certified random bits it consumes a previously existing random sequence of the order of $\sqrt{n}$ bits. The protocol of \cite{VV12} creates strings with $n$ random bits certified to be secure against quantum adversaries starting from a seed of a length of the order of $\log_2^3 n$ bits, offering an exponential expansion. 

Physically, the QRNG in \cite{PAM10} was implemented with trapped ion qubits \cite{OYM07} in order to close the detection loophole. Ion systems result in slower generation when compared to optical implementations, but offer almost perfect detection efficiency. Each atom first emits a photon with which it is entangled and then interference between the photons entangles the ions. This is a probabilistic heralded process. Experimental violation of Bell's inequality is a delicate task and the generation process was excruciatingly slow, giving only 42 certified random bits with a 99\% confidence level\footnote{The statistical nature of the device independent generation process can only certify a violation of Bell's inequality with a certain confidence level. We can ask for more certainty by taking more measurements (and thus reducing the generation speed).} after around a month of continuous running. 

Later proposals relax some of the requisites to allow for optical implementations and faster generation rates. Most optical detectors have a low efficiency, but transition-edge-sensor detectors \cite{LMN08} have been shown to offer a high enough efficiency to close the detection loophole in some modified versions of Bell's inequality \cite{GMR13} and have been used to generate certified quantum random numbers at a rate of about half a bit per second \cite{CMA13}.

The QRNG of \cite{CCG14} takes an alternative model that permits lower detection efficiencies with a semi-device-independent approach \cite{PB11} where we still do not trust the device but suppose we work with a quantum system with a bounded dimension. The experiment encodes the quantum data in the linear transverse momentum of single photons using spatial light modulators. While in the mentioned demonstration there are only two paths available, including spatial light modulators permits to control the spatial profile of single photons to encode higher dimensional quantum states. This optical system reaches bit rates of 0.28 certified bits per second. 

Other optical implementations focus on optimizing device independent random bit generation in experiments with entangled photon pairs. This is the approach in \cite{MSB15} and \cite{VSB15} and in the NIST randomness beacon \cite{NISTBeacon}.

The ideas of device independent quantum random number generators can be extended to an even more general model where quantum mechanics needs not to be true, following the example of the device independent quantum key distribution protocols \cite{BHK05,BCK12} that only require the no-signalling principle to hold. The no-signalling principle forbids the transmission of \emph{information} faster than the speed of light. A faster than light communication device would allow sending messages to the past and produces a conflict with causality \cite{Tol17}, as exemplified by the grandfather paradox\footnote{In the grandfather paradox, a time traveller, somewhat cruelly, decides to prevent the journey by killing his grandfather \cite{Nah99}. While it is still open whether General Relativity allows time travel, we can consider causality a fundamental principle. Even if it is not completely impossible, the no-signalling restriction is equivalent to asking an attacker for the highly nontrivial feat of time travel.}. The no-signalling principle is subtle. In entangled states, while there is non-locality and there are correlations that seem to travel faster than the speed of light, it is in fact impossible to use them to send information \cite{Die82,Bus82,Jor83}. 
 
In the device independent quantum random number generators of \cite{PAM10} and \cite{VV12} the bounds are also given for the non-signalling restriction. The exact bound on the conditional min-entropy changes, but the general results hold. In this new model, the protocols still work as randomness amplification schemes that need a uniform random seed. 

All the commented device independent random number generators, quantum and non-signalling alike, are, in fact, implementations of protocols that use the results from physical experiments to expand randomness. They start from a small random seed and produce a longer bit sequence guaranteed to be random. We give a more detailed description of this quantum randomness expansion in Section \ref{QRandExt}.

\subsection{Other forms of quantum certification}
Instead of testing locality with Bell inequalities, we can try to design certified quantum random number generators based on other experimental tests of the basic features of quantum theory. The Kochen-Specker theorem shows that there are states for which no non-contextual hidden variable model can reproduce the predictions of quantum mechanics \cite{KS67}. Contextuality in quantum mechanics is related to the existence of non-commutable observables where the order of measurement is important and there is no predefined model that can give the outcomes of two successive incompatible measurements. Contextuality implies nonlocality \cite{Ein48}. 

Quantum random number generators based on test of contextuality are designed to make sure we are accessing quantum randomness and not classical noise. In this model, we still work with untrusted devices but in a less adversarial setting. We assume the manufacturer of the random number generator is not actively trying to fool us, but we admit the device can be faulty or poorly designed. A test of contextuality shows whether we are truly reading bits from a quantum source or not. One of the advantages of quantum random number generators is that we can clearly trace the origin of our random bits to a defined quantum phenomenon. These certified generators can help to detect the randomness due to classical noise, imperfections or failures in the device and take only the randomness from quantum origin. Contextuality tests can work without spacelike separation of the devices. This is both the merit and the disadvantage of the method. These tests do not required complex nonlocal entangled states, but we cannot count on causality to guarantee the bits must be random. Unlike in device independent protocols, a rogue manufacturer can feed us pregenerated bits without being detected. 

The quantum random number generators of \cite{DZC13} and \cite{UZZ13} produce certified random bits based on contextuality tests through the violation of the Klyachko-Can-Binicioglu-Shumovsky, KCBS, inequality \cite{KCB08}, which doesn't require entangled states. The basic principle follows the model of \cite{PAM10}. A violation of the KCBS inequality guarantees a lower bound in the entropy of the output string, which can then be fed to a randomness extractor. The results serve as a certificate of quantumness, with a minimum amount of randomness that can be safely said to be of quantum origin. 

The physical implementation can be optical \cite{DZC13}, with a qutrit\footnote{The Kochen-Specker theorem works for any quantum system of dimension $d\geq 3$.} encoded in a photon in a superposition of three possible paths, or use a three-level trapped ion \cite{UZZ13}, which permits to close the detection efficiency loophole and avoids the problems of obtaining a single photon on demand. In the ion system, the random bits come from registering or not fluorescence during a measurement that takes around 10 ms. In both cases, under the tested experimental conditions, the devices could only provide a net gain in randomness, i.e. generate more random bits than they consumed, when using non-uniform measurement settings.

Along the same lines, there are also theoretical proposals for random number generators based on contextuality tests in settings similar to the previous experiment \cite{ACC12} and with entangled states \cite{ACS14} that highlight the relationship of randomness and incomputability \cite{CS08}.

\section{Postprocessing}
\label{postprocessing}
Standard random number generators are designed to produce a random uniform string. The postprocessing stage takes care of converting the raw bit sequence into a good quality output as close as possible to a uniform bit distribution. Postprocessing can include tasks like buffering to accumulate samples before generating the output strings or health tests that check the generator is working properly \cite{SK03}. For instance, the commercial quantum random number generator based on path branching Quantis includes hardware to check for inconsistencies following the AIS31 standard \cite{IDQ}.
 
Apart from these tasks, which vary from generator to generator, the main purpose of postprocessing is randomness extraction. Most physical RNGs include one form or another of randomness extraction to correct for biases and correlations that appear due to imperfections in the measurement and generation devices even for good randomness sources with a high entropy. 

A high entropy is not enough to guarantee the generated random sequence is fit for any purpose. While there are methods that can fix weak sources for their use in randomized algorithms \cite{Zuc96}, where randomness brings efficiency, not all protocols can work with imperfect randomness. In particular, many cryptographic protocols for tasks like bit commitment, encryption, zero knowledge or secret sharing are not secure unless they use an almost uniform random sequence \cite{DOP04}.

Some hardware random number generators mix different randomness sources by taking the logical XOR of their bits or feed the strings to a cryptographic hash function \cite{RFC4086}. Von Neumann proposed a simple debiasing method in which, for every pair of generated bits, we discard the results 00 and 11 and assign a 0 to 01 and a 1 to 10 \cite{vNe51}. If we have a systematic bias this method will remove it at the cost of throwing away at least half of our bits and reducing our bit rate at least by one fourth (discarding more bits the more biased our original sequence was). The basic method can be refined to improve its efficiency \cite{Eli72,Per92}. 

Before going on describing randomness extraction in more detail, it is important to define what is considered as an ``acceptably'' uniform output. A useful concept is that of distance between distributions. For two probability distributions $X$ and $Y$ defined in the same support (they can take the same values in a finite alphabet $\mathcal{A}$), we can define a statistical distance 
\begin{equation} 
d(X,Y)=\max_{a\in \mathcal{A}}|P_X(a)-P_Y(a)|.
\end{equation}
This metric gives the maximum difference in the probability of getting a particular result in the compared distributions. We say two distributions $X$ and $Y$ are $\epsilon$-close if
\begin{equation} 
d(X,Y) \leq \epsilon.
\end{equation}

In randomness extraction the goal is to produce an output sequence which is as close to uniform as possible. That usually means taking the $n$ bits of the raw output and transforming them into strings of $m$ bits with a distribution which is $\epsilon$-close to $U_m$ (a distribution uniform in $\{0,1\}^m$) for a small $\epsilon$ that depends on our requisites.

Ideally, we would like extractors that give as many output bits as possible with the smallest use of additional resources like computation time or additional randomness. In that respect, the randomness measures we have discussed in Section \ref{EntEstimation} serve as a design guide. In particular, the min-entropy of the distribution of the raw sequence gives a limit on how many bits we can extract. If we take $n$-bit strings from the raw sequence with a distribution $X$ of min-entropy $H_\infty(X)=k$, we can extract at most $k$ random bits that are close to uniform, irrespective of the original length. A random process is called an $(n,k)$-source if it produces $n$ bits with a distribution $X$ of min-entropy $H_{\infty}(X)\geq k$. 

In the following section we will discuss different methods to generate bit sequences as close to uniform as desired for rates close to the min-entropy limit and the advantages and limitations of different randomness extraction approaches.

\subsection{Randomness extractors}
Randomness extractors are functions that convert a weak source of entropy into a uniform bit generator. They were originally introduced in the study of randomized algorithms, but have become a basic tool in many areas of theoretical computer science. Randomness extractors and related concepts like dispersers, condensers and expander graphs have multiple applications and appear in the fields of pseudorandom number generators, error-correcting codes, samplers, expander graphs and hardness amplifiers, among others \cite{Vad07}. 

In this Section, we discuss only the few concepts about extractors most relevant to QRNGs and refer the interested reader to the extensive literature on the subject, ranging from introductory tutorials \cite{Sha11} to detailed surveys \cite{Nis96,NT99,Sha02}. There are many available options for randomness extraction and the final choice is usually influenced by the speed and hardware requirements of each method. Here, we just comment on some particularly interesting extractors.

In order to have an efficient method and preserve as many bits as necessary, we need to have a good estimation of our available entropy and then choose an adequate randomness extractor \cite{MXX13}. Otherwise, the output of the extraction function will not have the desired properties. 

In the following, we assume we have a well-characterized randomness source. The relevant entropy measures were discussed in Section \ref{EntEstimation}. The raw sequence is assumed to have a known min-entropy or, in some cases, at least some known properties such as independence between bits or that it comes from a Markov process.  

In the next Sections, we also assume by default that we want an $(n,m,k,\epsilon)$-extractor: a function that convert $n$ bits of an $(n,k)$-source into $m$ output bits with a distribution that is $\epsilon$-close to uniform, with $m$ as close to $k$ as possible.

\subsubsection{Deterministic extractors}
\label{DetExt}
Deterministic extractors are functions 
\begin{equation}
\label{defnm}
\text{Ext}: \{0,1\}^n \rightarrow \{0,1\}^m
\end{equation}
that take input strings of $n$ bits $\{0,1\}^n$ into $m$ output bits. They are particularly attractive as they are deterministic algorithms that only need an input sequence to work. However, they have some limitations that prevent their use with certain randomness sources. 

As in all extractors, we can only produce an output close to uniform if the input sequence already has enough intrinsic entropy. If the input sequence is an $(n,k)$-source, a necessary condition for the output sequence to be close to uniform is that $m\geq k$. Unfortunately, the necessary condition is not sufficient and we can only find deterministic extractors for certain limited input distributions. 

An elementary argument shows the impossibility of general deterministic extractors. Imagine a function from $\{0,1\}^n$ to $\{0,1\}$. We can divide all possible inputs into one set of all the input $n$-bit strings that give a $0$, $\text{Ext}^{-1}(0)$, and another set that is taken to $1$, $\text{Ext}^{-1}(1)$, and at least one of them has a size $2^{n-1}$ or larger. An input that is a uniform distribution in the larger set has at least min-entropy $n-1$ but produces always the same output showing there is no one-size-fits-all extractor valid for any input distribution \cite{CG88}.

There are, however, valid extractors for input distributions belonging to certain families of processes that describe reasonable sources. Among others, there are practical deterministic extractors for samplable distributions \cite{TV00}, for bit-fixing sources where an adversary can set part of the bits \cite{GRS06,KZ07} and generalizations for affine sources \cite{GR05,Bou07} or sources with an output that is distributed uniformly over an unknown algebraic variety \cite{Dvi12}. 

Variable length deterministic extractors form another group of interesting deterministic extractors which deviate slightly from the description of Equation (\ref{defnm}). They are exemplified in the von Neumann algorithm: a deterministic method that works for an unknown distribution and gives an output of a length that is not known before the extraction. In the von Neumann randomness extractor described at the beginning of this section the only requisite is that each input bit is independent from the previous and following bits. Refined versions of von Neumann's method reduce the discarded entropy and give efficiencies close the information theory limit given by the Shannon entropy of the source \cite{Eli72,Per92}. Further modifications give algorithms that produce unbiased sequences on the more general condition that the input sequence comes from a Markov chain \cite{Blu86,ZB12}.

The main appeal of the original method is its simplicity. It requires minimal computing power, it can be implemented with just basic hardware and the distribution at the source needs not to be perfectly known. However, it has some important limitations. If we have an external attacker that can alter the bias from bit to bit, even slightly, the von Neumann extractor no longer works. In fact, there is no deterministic algorithm that can give a uniform output for a random variable $X=(X_1,X_2,\ldots, X_n)$ with $n$ bits if the bias of the input bits can vary so that the probability of finding a $1$ for the $n$th bit conditioned on the measured string for the previous bit values $s$ is 
\begin{equation}
\label{SV}
\delta\leq  P_{X_{i}}(1|x_1 x_2\ldots x_{n-1}=s)  \leq 1-\delta
\end{equation}
for a $0<\delta\leq \frac{1}{2}$. This is called a Santha-Vazirani source and was described as a model for weak randomness sources in \cite{SV86} together with an impossibility proof for a deterministic extractor. 

Despite this limitation, there are deterministic algorithms that permit to use a weak Santha-Vazirani source to simulate randomized algorithms \cite{VV85b,ACR99}. The requisites for randomization are less stringent than for other applications, like cryptography, and weak sources that fail to produce nearly uniform outputs are sometimes valid.

Even if we use a deterministic extractor, a single weak source is not good enough for many cryptographic protocols. While weak randomness can be used securely with signature schemes, encryption and other related protocols need a high quality key or they become vulnerable \cite{MP91,DS02,DOP04,ACM14}. 

For applications where we need an output close to uniform, \cite{SV86} offer a simple solution: combining the output of two independent Santha-Vazirani weak sources we can produce output sequences that cannot be distinguished by any polynomial-time algorithm from a uniform distribution.  As long as we have access to a physical method that produces \emph{some} randomness, we can generate bit strings that cannot be distinguished from a random string with any efficient algorithm. This is just as good as true randomness for the vast majority of applications of randomness, including cryptography. 
 
{\bf Multiple source extractors} follow this model and take the output of two or more weak sources and process them to generate a sequence that is close to uniform. There are many methods that depend on the concrete input distributions, the number of sources we have and the desired properties of the output sequence. 

A simple extractor valid for two $n$-bit blocks from two independent weak sources, both with min-entropy at least $n/2$, is taking the $G\!F(2)$ inner product of the $n$-bit blocks, which reduces to computing the parity of the bitwise AND of the two sequences \cite{CG88,Vaz87,Vaz87b}. 

Other representative methods to combine different randomness sources can be found in \cite{DEO04,Bou05,Raz05,BIW06,Sha08,Rao09}. 

The idea of combining sources is also behind the second main group of randomness extractors, seeded extractors. We can consider them a special case of multiple source extractors with one weak source and a perfectly uniform source that only produces a small amount of bits. 

\subsubsection{Seeded extractors}
\label{SeedExt}
As we have seen, for many raw bit distributions, we can only achieve an output close to uniform with the help of some additional randomness. In seeded extractors we have a function
\begin{equation}
\text{Ext}: \{0,1\}^n \times \{0,1\}^d \rightarrow \{0,1\}^m
\end{equation}
that takes as its input $n$ bits from the raw sequence and a uniform random seed of $d$ bits to produce $m$ output bits. We assume $d$ is much smaller that $m$. With the addition of the seed, which plays a role similar to the seed in pseudorandom number generators, we can guarantee that there exist extractors that produce an almost uniform output close to the maximum possible length. We call a $(k,\epsilon)$ extractor to a function that, for any input $k$ source (a raw sequence of, at least, min-entropy $k$), produces an output sequence that is $\epsilon$ close to uniform. The seed acts as a catalyst that permits to find general methods that will always work. 

Seeded randomness extractors were first defined in \cite{NZ96} in the context of randomized algorithms. Using the probabilistic method \cite{AS06}, \cite{RT00} showed there always exist extractors with an output that contains almost all of the available hidden entropy in an input raw sequence coming from any $k$-source. For input blocks of $n$ bits from a $k$-source, we can build extractors with an output of a size $m\approx k +d $ that is $\epsilon$-close to uniform using only a seed of a length $d$ of the order of $\log_2(n)$. There are different explicit constructions for these seeded extractors, like the ones in \cite{TaS96,LRV03}.

The need for a uniform seed seems a contradiction: we require the resource we are trying to produce. However, the requisites on the seed are less restrictive than it seems. In many explicit extractors the seed has a length logarithmic in the size of the input string. For a small enough $d$, we can even replace the requisite of randomness by an exhaustive enumeration of all the $2^d$ possible sequences. In randomized algorithms, enumeration followed by majority voting permits to simulate a good uniform source \cite{GW02}. However, this approach is clearly not valid for cryptography, where we need unpredictability.

In quantum random number generators, seeded extractors provide protection against external attackers. There are constructions for which there exist proofs of security against quantum attackers of different power \cite{BT12}.

A first notable result is the Trevisan extractor \cite{Tre01}, an explicit construction which has some nice properties like its resistance against quantum adversaries \cite{DV10,DPV12,TaS11} and the way it preserves the randomness of its seed \cite{MPS12}. The Trevisan extractor is built on the Nisan-Widgerson pseudorandom number generator \cite{NW94}. It can be seen as a random function whose truth table is given by the bits from the weak source. The random function expands the $d$ bits of a uniform random seed, both in the PRNG and the extractor sense. Different variations of the Trevisan extractor have been implemented for their use with quantum random number generators and in quantum key distribution \cite{MPS12,MXX13}. Their main advantage is that the size of the random uniform seed is only poly-logarithmic in the size of the input blocks. However, practical implementations can slow down the bit generation process due to the involved calculations required during the extraction.

A second general method of particular interest is two-universal hashing. The Leftover Hash Lemma \cite{ILL89,HIL99} shows that the output of a two-universal hash function with an input with high enough entropy is almost uniformly random. Two-universal hash functions, such as the families introduced in \cite{CW79,WC81}, can extract the randomness in a weak source in a secure way in the presence of an eavesdropper. If we have a good estimation or a conservative bound on the correlation of our weak random source with the eavesdropper, using the conditional entropies described in Section \ref{EntEstimation}, it is possible to use a generalization of the Leftover Hash Lemma with side information \cite{TSS11}. In the most general case, the side information can also be quantum. In a quantum random number generator with technical noise, we can assume that all the randomness that comes from imperfections or otherwise does not adjust to our model of the quantum system that produces the raw bits is due to an eavesdropper. In those conditions it is still possible to design a seeded extractor that gives an almost uniform output that is independent from external systems \cite{KT08,KR11}. These methods are also applied in privacy amplification in Quantum Key Distribution \cite{BBR88,BBC95,RK05,KMR05}. 

Randomness extraction with two-universal or, more generally, $l$-universal hashing forces us to use a relatively long seed, comparable to the size of the block $n$, but it can be recycled. A randomly chosen public uniform seed can be reused and permits a secure seeded extractor in the presence of an imperfect randomness source under partial influence of an attacker \cite{BST03,Sko15}.

When compared to implementations of the Trevisan extractor, this method offers a fast extractor function that takes less computational resources at the cost of a larger seed \cite{MXX13}. Some implementations, like hashing with Toeplitz random binary matrices \cite{MNT90,Kra94}, are particularly efficient. We can define one such extractor where the seed is used as a rectangular matrix that is multiplied to $n$-vectors from the source to produce an output of almost independent bits \cite{FRT13}. This approach is used in some commercial devices which include the extraction function as a precomputed random matrix that acts as the seed and is distributed coded into the device \cite{TR12}. While ensuring the seed is uniformly random to a high degree is a painstaking task, it only needs to be done once. Long unsophisticated methods, like repeatedly taking the XOR of multiple independent generators, are acceptable.

\section{Quantum randomness extractors: randomness expansion and randomness amplification}
\label{QRandExt}
Quantum mechanics does not only offer new sources of entropy for random number generators, but also new protocols related to randomness extraction. We will consider physical randomness extractors which use untrusted ancillary systems either to expand the random output of a uniform source or to turn a weak randomness source into strong one \cite{CSW14}. 

There are two interesting families of protocols: quantum randomness expansion and quantum randomness amplification. In quantum randomness expansion, we start from small random seed and, with the help of a quantum protocol, we produce a longer bit sequence with strong guarantees of randomness. In randomness amplification we take a weak source, either classical or quantum, and use a quantum system to amplify the randomness in the weak source and give an arbitrarily close to uniform output. 

Related to these ideas is also privacy amplification, where we take a bit string which is partially known to an adversary and produce a smaller sequence for which no external attacker can have any statistically significant information. There are known classical \cite{BBR88,BBC95} and quantum \cite{DEJ96} algorithms for this task, but we can also use methods related to randomness extraction protocols that can guarantee the output is uncorrelated to any causally preceding events and, therefore, must be private.

In this Section, we given an overview of the main ideas behind these concepts. The reader can also find a good review of all the mathematics involved in \cite{PP15}.

\subsection{Quantum randomness expansion}
Quantum randomness expansion protocols follow the model of seeded randomness extractors (see Section \ref{SeedExt}): assisted by a random seed, we process the bits from a weak randomness source and give an output that is as close to uniform as desired.

All the device independent generators discussed in Section \ref{DevInd} are, indeed, implementations of some kind of randomness expansion protocol working on the weak randomness produced in the nonlocality experiments of different Bell tests. The quantum system serves both as a weak source of randomness and as a way to guarantee the privacy of the results. The random seed serves as a starting point to take the randomness in the quantum devices into a uniform output. 

Randomness expansion protocols can be concatenated using a limited number of devices \cite{MS14}. By repetition of simple protocols with a finite number of quantum devices, we can increase the size of the output arbitrarily to produce sequences certified against quantum adversaries \cite{CY14}. 

If we relax our requirements and trust part of the system, we can also find semi-device independent randomness expansion protocols. For instance, for unstrusted devices but a trusted quantum state with a bounded dimension, the protocol in \cite{BPP14} gives an expansion scheme that does not require entanglement, which makes it easier to implement in practice. If we consider an adversary which does not directly control our device, but can characterize it better than us and has a complete model of its inner workings, we can also produce a private output string if we make full use of all the data taken from a series of Bell tests instead of restricting to the usual inequalities \cite{BSS14}. 

A different kind of extractor without Bell tests is the source-independent seeded extractor in \cite{CZY16}, which is designed to work with imperfect quantum sources and addresses many problems of optical quantum random number generators like losses, multiphoton pulses or unbalanced beam splitters. 

Similarly, there are also quantum-to-classical randomness extractors that give a procedure to measure a quantum state from a source that can be correlated to an eavesdropper so that we maximize the amount of random bits we get without giving away information to the adversary \cite{BFW14}.

Finally, the concepts of randomness expansion can also be formulated as a privacy amplification problem in which we want to extend the length of a private string while keeping it secret under the usual assumptions of the device independence scenario with untrusted equipment \cite{CK11}. The task is possible and efficient against quantum attackers, but, unlike other protocols, there are severe limitations if we consider attackers that are only restricted by nonsignalling constraints \cite{AT12}. Anyway, while considering nonsignalling attackers gives quite general security results, quantum mechanics seems to be the nonlocal theory that best describes the physical world and a quantum secure protocol can be safely considered as valid.

\subsection{Quantum randomness amplification}
The need for a uniform seed in device-independent protocols comes from two parts of the procedure. First, in Bell tests we assume we have uniform random bits to choose the measurement settings. Second, the generated bit sequence is only guaranteed to have a lower bound on min-entropy, but we need to use some seeded randomness extractor to obtain a uniform output bit string.

Quantum randomness amplification protocols eliminate these previous uniform randomness requisites and give a way to use a weak source in combination with quantum devices to produce uniform random bits. In Section \ref{DetExt} we have seen it is impossible to find a general deterministic method to extract randomness from any limited min-entropy source, even from restricted weak origins of entropy like Santha-Vazirani sources. With the help of quantum mechanics, we can solve this problem and find methods to extract almost uniform randomness in those situations. From a certain point of view, these protocols are not so much deterministic randomness extractors as multiple source extractors where we prove how to combine the randomness in the quantum devices with the randomness of a weak source to produce a good quality output. While the exact details vary from protocol to protocol, the quantum part is usually limited to simple measurements on the different subsystems of an entangled state. From an experimental point of view, the hardest requisite to satisfy is making sure the quantum devices are independent, which can be a problem in protocols that require multiple devices.

A remarkable contribution to quantum randomness amplification is the randomness amplification protocol of \cite{CR12}, which shows there are deterministic protocols that can amplify the randomness in Santha-Vazirani sources using ancillary physical systems. The result rests only on nonlocality and is robust against attackers that can go beyond quantum mechanics. This protocol needs a large supply of imperfect randomness. One natural application would be using quantum randomness amplification only to provide the random seed for the quantum randomness expansion protocols of the previous Section and then use the less involved quantum randomness expansion protocols to generate the final random bit stream.

While the original protocol works only for small biases $\delta$ in the definition of the source, see Eq. (\ref{SV}), \cite{GMT13} give a quantum randomness amplification protocol that is valid for arbitrarily weak sources of entropy. Further protocols can take any input weak source with a bounded nonzero min-entropy \cite{BPP14,CSW14,PP14} and give practical ways to use Santha-Vazirani sources, requiring only a limited number of independent devices \cite{BRG16}.

There are also interesting ramifications for fundamental science experiments. Many of the concepts of quantum randomness amplification can be traced back to the study of randomness in Bell inequalities. These results are interesting in themselves as they determine which random number generators can be used in the foundational experiments on nonlocality in Bell tests. In Bell experiments there is a ``free will'' loophole: if the settings in the measurement are correlated, the violation of a Bell inequality cannot be used as a guarantee against an eavesdropper \cite{KHS12}. Fortunately, even in the usual experiments, there is a certain tolerance for small correlations \cite{Hal10}, but general min-entropy sources are not valid for the selection of the settings in Bell experiments \cite{TSS13}.

\section{Randomness testing}
\label{tests}
Once we have generated a raw random sequence, we need to do some quality checks to be sure the device is working correctly. Unfortunately, there is no way to check a finite sequence is truly random. Taken to its most absurd extreme, it is like asking whether a 0 bit is fundamentally more random than a 1. Apart from the uncomputable Kolmogorov complexity \cite{LV08}, there is no way to deduce that a random string is really random, but there are methods to detect suspicious sequences. While the bit string 1111111111 is just as likely as 0100110111, if we have a generator that consistently outputs more ones than zeros we have reason to suspect it is not acting randomly. 

The customary approach to randomness testing is using a series of statistical tests. Knuth covers some of the most usual ones in \cite{Knu97}. The main suites available to perform these statistical tests are the NIST \cite{NIST10}, TestU01 \cite{LS07} and the DieHard and DieHarder \cite{Mar96,DieHarder} suites. There are also special-purpose randomness testing batteries, like the one included with the SPRNG software \cite{SMC03}, which is designed to check for problems in parallel implementations of pseudorandom number generators. 

These suites include different tests. In the following list, we present some of the most relevant tests to give a feeling of the kind of hidden correlations that can appear. 

\begin{enumerate}
\item The \emph{frequency (monobit)} test, which calculates the proportion between ones and zeroes and how close that proportion is to $\frac{1}{2}$, and frequency tests within a block, similar to the previous one, but testing for the expected probabilities for the specified block sizes.
\item The \emph{runs} test, which checks if the number of \textit{runs}\footnote{A \textit{run} is defined as an uninterrupted sequence of identical bits bounded by a bit of the opposite value before and after the same-bit sequence.} in a bit string corresponds to that in a random sequence and if the oscillation between zeroes and ones is too fast or too slow. 
\item The \emph{spectral} test, which tries to detect periodic features in the sequence that would indicate a deviation from the assumption of randomness. 
\item Maurer's \emph{Universal Statistical} test \cite{Mau92}, which detects whether or not the sequence can be significantly compressed without loss of information.
\item \emph{Autocorrelation} tests which check the correlation of the sequence with shifted versions of itself.
\end{enumerate}
Most tests apply statistical analyses similar to the standard chi-squared test. The result is a \textit {p-value} that indicates how likely it is for a purely random number generator to produce the tested sequence. Each test suite has different threshold values to determine if a given p-value is compatible with randomness or not. 

These tests, while useful to detect faulty generators, cannot prove a generator produces truly random outputs. Deterministic pseudorandom number generators like the Mersenne Twister can pass the tests but are predictable. Likewise, there can be false positives for correlations and the tests should be run multiple times for each generator. Statistically, even a perfect random number generator would fail a test from time to time.

Testing is also vulnerable to an active attacker that feeds us pregenerated random sequences that pass the tests. In Section \ref{DevInd} we have described some quantum protocols to solve this issue. 

Apart from that, the tests are usually designed with pseudorandom number generators in mind and do not include physical models into account. Some correlations due to implementation-related problems, like afterpulsing in photon detectors, are not specifically checked. 

All these problems notwithstanding, any good quantum random number generator should be able to pass all the tests in any given suite and using some form of randomness testing during operation can help to detect sudden failures or faulty components. 

\section{Discussion}
\label{discussion}
Quantum random number generation is probably the most mature quantum technology. We have seen the multiple ways we can harness the randomness in quantum mechanics to produce random bit strings. Physical phenomena such as radioactive decay, photon splitting, noise in Raman amplification, laser phase noise or amplified spontaneous emission can serve as reliable entropy sources. 

We have reached a point where optical quantum random number generators routinely reach generation rates in the order of megabits per second with promises of gigabit rates and new generation methods are still being suggested every year. While there is a race to announce the highest possible generation rates, in many cases, the actual implementation is limited by practical hurdles in the speed of the electronic systems and the postprocessing methods. 

Many proposals focus on the generation principle, on making sure the quantum phenomenon of interest produces fresh entropy at a fast rate, but do not deal with making full use of the available bits and give random bit rates which are only true as an extrapolation. In the research phase, it is perfectly acceptable to leave all the processing details for later and work on a limited collection of stored samples, but, at this point of development, there is a need for better and faster production of the final, usable random bits.

Commercial devices, by necessity, have these aspects covered but they still offer bit rates with a gap around two orders of magnitude with respect to the fastest possible lab rates. In some applications, like simulation, this is important, as quantum random number generators have to compete against fast pseudorandom number generators that work essentially at the speed of the available processor. 

Concerning the bit rate, there are two relevant issues. One is the communication bottleneck. External devices will always need a communication channel with the computer that uses the random bits. The fastest USB protocols (USB 3.0 and 3.1) and PCI Express components can reach communication rates in the order of tens of Gbps that is enough for many generators. Alternatively, many optical implementations can be adapted or have been demonstrated to work in integrated silicon setups that could be included as part of future processors. 

Communication at those rates is challenging, but it is an engineering problem that can be solved with current technology with the right systems. A second more interesting limitation is randomness extraction. In Section \ref{postprocessing}, we have described different ways to turn the raw bits coming from measurement and the first simple conditioning into good quality random bits. While some quantum random number generators are claimed to directly produce random enough raw sequences, in some applications like cryptography, less than perfect uniformity can pose serious problems. In general, quantum random number generators should include a well-designed postprocessing phase.

Seeded extractors like Trevisan's or two-universal hashing have good security properties against quantum attackers. That should be the standard that postprocessing methods should aspire to. At the moment, postprocessing is relatively slow when compared to the potential generation rates of the fastest optical generators. The most efficient implementations use postprocessing based on two-universal hashing with binary matrix multiplication. There is a large open area of research on identifying and constructing new extractors that are resistant against quantum attacks and can be fast enough to sustain output bit rates in the order of Gbps. 

Self-testing is another area for future improvement. Physical random number generators can fail due to component degradation or even external attacks. In Section \ref{DeviceIndependent} we have described many possible approaches to quality control. In particular, device independent protocols offer reliable random numbers even if we don't trust our hardware. Device independent randomness generation and quantum randomness expansion and amplification are quite active areas of research and the last years have seen many interesting results, including new protocols based on nonlocality that can perform classically impossible tasks, like physically-assisted deterministic randomness extraction from weak sources. 

Device independent quantum random number generators are still experimentally challenging and produce bits at sluggish rates. In Section \ref{DeviceIndependent} we have also commented on more relaxed approaches to certification, but this is likely to be an active area for the next years, both in technological development research to make better device independent QRNGs and in the theoretical search for simpler paths to certification.

At the moment of writing, both pure and applied research have reached an interesting point where there are new fundamental results and, at the same time, there appear different quantum random number generators in the market. 

With this review, we hope we have introduced the reader to the existing technologies and hinted at some future directions. 

\begin{acknowledgments}
This work has been funded by Project TEC2015-69665-R (MINECO/FEDER, UE).
\end{acknowledgments}
%

\end{document}